\documentclass[final,onecolumn,prf,aps,amsfonts,amssymb,showpacs,superscriptaddress,floatfix,nofootinbib]{revtex4-1}
\usepackage{amsmath}
\usepackage{graphicx}
\usepackage[dvipsnames]{color}
\usepackage{hyperref}
\usepackage{psfrag}
\usepackage{stmaryrd}
\usepackage{amssymb}
\usepackage{wasysym}
\usepackage{float}
\usepackage{color}

\usepackage[T1]{fontenc}
\usepackage[french,english]{babel}
\usepackage[utf8]{inputenc}
\usepackage{lmodern}

\usepackage{geometry}
\begin{document}

\title{Three-dimensionality of the triadic resonance instability of a plane 
inertial wave}

\author{Daniel Odens Mora}
\affiliation{Universit\'e Paris-Saclay, CNRS, FAST, 91405 Orsay, France}
\author{Eduardo Monsalve}
\affiliation{Universit\'e Paris-Saclay, CNRS, FAST, 91405 Orsay, France}
\author{Maxime Brunet}
\affiliation{Universit\'e Paris-Saclay, CNRS, FAST, 91405 Orsay, France}
\author{Thierry Dauxois}
\affiliation{Universit\'e de Lyon, ENS de Lyon, CNRS,
    Laboratoire de Physique, F-69342 Lyon, France}
\author{Pierre-Philippe~Cortet}
\email[]{pierre-philippe.cortet@universite-paris-saclay.fr}
\affiliation{Universit\'e Paris-Saclay, CNRS, FAST, 91405 Orsay, France}

\date{\today}

\begin{abstract}
We analyze theoretically and experimentally the triadic 
resonance instability (TRI) of a plane inertial wave in a rotating fluid. 
Building on the classical triadic interaction equations between helical modes, 
we show by numerical integration that the maximum growth rate of the TRI is 
found for secondary waves that do not propagate in the same vertical 
plane as the primary wave (the rotation axis is parallel to the vertical). In 
the 
inviscid limit, we prove this result analytically, in which case the change in 
the horizontal propagation direction 
induced by the TRI evolves from $60^\circ$ to $90^\circ$ depending on the 
frequency of the primary wave. Thanks to a wave generator with a large spatial 
extension in the horizontal direction of invariance of the forced wave, we 
are able to report experimental evidence that the TRI of a plane inertial 
wave 
is three-dimensional. The wavevectors of the secondary 
waves produced by the TRI are shown to match the theoretical predictions based 
on the maximum growth rate criterion. These results reveal that the triadic 
resonant interactions between inertial waves are very efficient at 
redistributing energy in the horizontal plane, normal to the rotation axis. 
\end{abstract}

\maketitle

\section{Introduction}

Rotating and stratified fluids allow the propagation of waves in their bulk, as 
a result of the restoring action of the Coriolis force and of the buoyancy 
force, respectively~\cite{Greenspan1968,Lighthill1978,Sutherland2010}. 
Moreover, inertial waves in rotating fluids and internal gravity waves in 
stratified fluids share several remarkable features: they have similar 
dispersion relations linking the ratio between the wave frequency and the 
rotation rate or the buoyancy frequency to the tilt angle with the horizontal 
of the direction along which their energy propagates (with the rotation or 
gravity axis parallel to the vertical). As a consequence, their group and phase 
velocities are normal to each other. Also, their wavelength is independent of 
their frequency and is set by boundary conditions, viscous dissipation and 
non-linearities~\cite{Brunet2019}. This leads to a variety of wave structures 
like self-similar 
beams~\cite{Mowbray1967,Thomas1972,Flynn2003,Cortet2010,Machicoane2015}, plane 
waves~\cite{Mercier2010,Bordes2012,Bourget2013}, resonant cavity 
modes~\cite{Aldridge1969,McEwan1970,Maas2003b,Boisson2012,Boisson2012b} and 
even cavity limit cycles called wave 
attractors~\cite{Maas1997,Rieutord2001,Manders2003,Grisouard2008,Klein2014,Brunet2019}.

Global rotation and density stratification are two major ingredients of atmospheric and oceanic turbulent dynamics~\cite{Pedlosky1987}. Inertial and internal gravity waves are therefore important players in these geophysical flows in which they merge into inertia-gravity waves with a dispersion relation coupling rotation and buoyancy~\cite{Lighthill1978,Pedlosky1987}. In this context, Wave Turbulence Theory (WTT), which addresses the statistical properties of weakly nonlinear ensembles of waves in large domains~\cite{Zakharov1992,Newell2011,Nazarenko2011}, stands as an interesting direction for improving turbulence parametrizations in coarse atmospheric and oceanic models~\cite{Gregg2018}. This is particularly the case since several recent studies have given credence to the WTT framework for inertial waves in experiments~\cite{Monsalve2020} and in numerical simulations~\cite{Yokoyama2020,LeReun2021} as well as for internal gravity waves in experiments~\cite{Savaro2020,Davis2020}.

In the framework of WTT, an energy cascade towards small scales and small 
frequencies emerges as the statistical result of weakly nonlinear interactions 
within resonant triads of 
waves~\cite{Galtier2003,Cambon2004,Lvov2004,Nazarenko2011b}. A fundamental 
process at play in this weakly non-linear cascade~\cite{Smith1999} is the triadic resonance 
instability (TRI) which drains the energy of a primary wave at frequency 
$\sigma_0$ toward two subharmonic waves at frequencies $\sigma_1$ and 
$\sigma_2$ such that $\sigma_1+\sigma_2=\sigma_0$. The instability of inertial 
and internal gravity waves has been reported for a long time with early 
works in the 1960's (see \cite{Staquet2002} and references therein). Several 
quantitative experimental and numerical studies of the TRI have been conducted 
since the 2000's, starting with the 2D numerical simulations of a propagating 
plane internal gravity wave by Koudella and Staquet~\cite{Koudella2006}. Since 
then, the TRI has been characterized numerically and experimentally for plane 
waves~\cite{Bordes2012,Joubaud2012,Bourget2013,Bourget2014} and for the 
self-similar beam of wave 
attractors~\cite{Jouve2014,Scolan2013,Brouzet2017,Brunet2019}. Besides, 
refinements of the theory for the TRI accounting for finite size effects, i.e.,
the finite number of wavelengths present in the primary wave beam, have been 
proposed~\cite{Bourget2014,Karimi2014}.

In all these works, when the comparison of the experimental or numerical data 
with the theoretical framework of the TRI is done, it is restricted to the case 
where the secondary waves propagate in the same vertical plane as the primary 
wave, assuming that the secondary waves are invariant in the same 
horizontal direction as the primary wave (labeled direction $y$ in the 
following). This implicit assumption of a two-dimensional instability is somewhat consistent with the considered numerical and experimental setups. For 
example, in the experiments of 
Refs.~\cite{Bordes2012,Joubaud2012,Bourget2013,Bourget2014,Scolan2013,Brouzet2017},
 with the notable exception of the work of Brunet~\textit{et 
al.}~\cite{Brunet2019}, the width of the primary wave beam in the $y$-direction was 
neither large compared to its wavelength nor to its typical propagation 
distance. Furthermore, in the 2D numerical simulations of 
Refs.~\cite{Koudella2006,Jouve2014} the flow was strictly invariant in the 
$y$-direction. On the one hand, the possibility for the triadic resonance 
instability to be three-dimensional, i.e., with an energy transfer toward two 
waves propagating in other vertical planes than the one of the primary wave, 
has yet to be 
considered theoretically. On the other hand, this possibility has not been 
tested either because of the very limited extension of the forcing in the 
$y$-direction in experiments or because 2D simulations render it forbidden at 
the outset.

In the present article, we analyze theoretically and experimentally the 
triadic resonance instability of a plane inertial wave in a rotating fluid of 
uniform density. We first show by numerical 
integration that the classical triadic resonance interaction analysis for the 
TRI of a plane inertial wave predicts a maximum growth rate for secondary waves 
propagating out of the primary wave plane. We moreover analytically demonstrate 
this result in the inviscid limit. Second, we test this theoretical prediction 
experimentally by forcing a plane inertial wave beam with an extension 
in its horizontal invariance direction $y$ much larger than its wavelength. We 
find a good agreement between the features of the secondary waves produced by 
the instability in the experiments and the predictions for the wave triad 
maximizing the TRI theoretical growth rate. Thus, we confirm the natural 
tendency of the TRI of a plane inertial wave to be three dimensional and to 
redistribute the energy in the horizontal plane normal to the rotation axis.

\section{Triadic resonance instability of a plane inertial wave}\label{sec:theory}

\subsection{Navier-Stokes equation in a rotating frame}

In the following, we consider the dynamics of a fluid of uniform density 
subject to a global rotation at a rate 
$\Omega$ around the vertical axis defined by the unit vector $\hat{\bf z}$.
In the rotating frame of reference, the velocity field ${\bf u}({\bf x},t)$ of 
incompressible fluid motions ($\nabla \cdot {\bf u} = 0$) is described by the 
Navier-Stokes equation
\begin{eqnarray}\label{eq:nsrot}
    \frac{\partial {\bf u}}{\partial t}+({\bf u}\cdot\nabla){\bf 
    u}&=&-\frac{1}{\rho}\nabla p - 2 \boldsymbol{\Omega}\times {\bf u} + 
    \nu\nabla^2 {\bf u},    
\end{eqnarray}
where $p$ is the pressure field, $\nu$ the fluid viscosity, $\rho$ the fluid density and $\boldsymbol{\Omega}=\Omega \, \hat{\bf z}$ the vector rotation rate. In the inviscid and linear limit, Eq.~(\ref{eq:nsrot}) has anisotropic, dispersive and helical plane wave solutions, called inertial waves \cite{Greenspan1968,Pedlosky1987}. Their dispersion relation,
\begin{eqnarray} \label{eq:dispersion}
\sigma= s \, 2\Omega\, \frac{{\bf k}\cdot 
\hat{\bf z}}{|{\bf k}|} \, ,
\end{eqnarray}
relates the normalized wave angular frequency $\sigma^*=\sigma/2\Omega$ to the 
direction of the wavevector ${\bf k}$. In Eq.~(\ref{eq:dispersion}), $-s$ is 
the sign of the wave helicity $(\nabla \times {\bf u}) \cdot {\bf u}$~\cite{Smith1999}.
The dispersion relation~(\ref{eq:dispersion}) reveals that the 
wavelength $\lambda=2\pi/k$ (where $k=|{\bf k}|$) is independent of the 
frequency $\sigma$. In practice, the wavelength is set by the boundary 
conditions, 
viscous effects, and even by non-linear effects in some cases 
(see~\cite{Brunet2019} for a discussion on this point). When viscosity is 
considered in Eq.~(\ref{eq:nsrot}), the amplitude of the wave of wavevector 
${\bf k}$ is damped at a rate $\nu k^2$~\cite{Machicoane2018}. Besides, viscosity does not modify the 
wave dispersion relation~(\ref{eq:dispersion}) (see Ref.~\cite{Machicoane2018}) which is not the case, e.g., for internal gravity 
waves. Finally, it is worth mentioning that inertial plane waves are also 
solutions of the complete (non-linear) Navier-Stokes equation~(\ref{eq:nsrot}), 
in which case they are however not necessarily stable as we will see in the 
following.

\begin{figure}
	\centerline{\includegraphics[width=14cm]{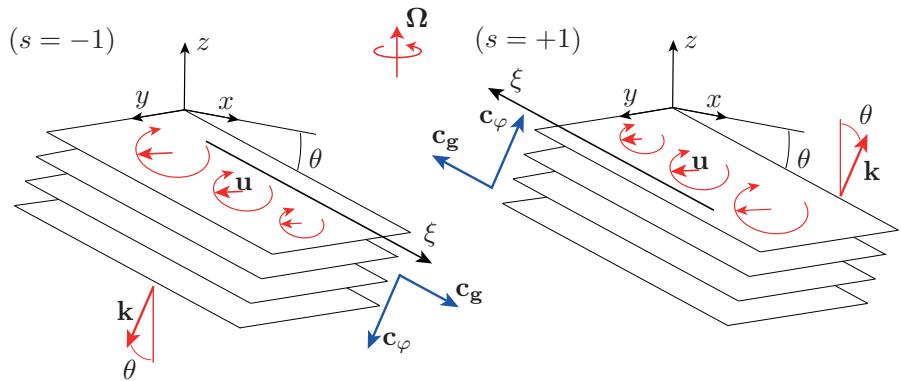}}
	\caption{Sketch of a plane inertial wave of wavevector ${\bf k}$ with a 
	polarity $s=-1$ (left) and $s=+1$ (right). The wave is invariant in the 
	horizontal $y$-direction ($k_y=0$). The fluid motions consist in an 
	anticyclonic circular translation in the planes of constant phase, normal 
	to ${\bf k}$, which are tilted by an angle 
	$\theta=\cos^{-1}(\sigma^*)=\cos^{-1}(k_z/k)$ with respect to the 
	horizontal. The phase of the wave propagates normally to these constant 
	phase planes, but the energy of the wave propagates parallel to these 
	planes along the group velocity. The vectors ${\bf c_g}$ and ${\bf 
	c_\varphi}$ 
	indicate the direction of the group and phase velocities, respectively. The 
	amplitude of the fluid motions is damped along the energy propagation 
	direction ${\bf c_g}$ at a rate $\nu k^2$.}\label{fig:plane_wave}
\end{figure}

The structure of a plane inertial wave of wavevector ${\bf k}$ is sketched in 
Fig.~\ref{fig:plane_wave}, for the two possible polarities $s=-1$ and $s=+1$. 
The fluid motions in the wave consist in an anticyclonic circular translation 
at frequency $\sigma=s 2\Omega\, k_z/k$ in the planes of constant phase, which 
are normal to ${\bf k}$ and therefore tilted by an angle 
$\theta=\cos^{-1}(\sigma^*)=\cos^{-1}(k_z/k)$ with respect to the horizontal 
($k_z={\bf k}\cdot \hat{\bf z}$). The phase shift of the motion between close 
parallel planes of constant phase involves a shear and finally leads to a 
vorticity $\nabla \times {\bf u}$ that is parallel to the local velocity ${\bf 
u}$. The energy of the wave propagates parallel to the slope of the planes of 
constant phase at the group velocity $|{\bf c_g}|=2\Omega \sin \theta/k$. The 
energy goes upwards (with respect to $\hat{\bf z}$) when $s=+1$ and downwards 
when $s=-1$. The phase of the wave propagates at the phase velocity ${\bf 
c_\varphi}=\sigma {\bf k}/k^2$ which is normal to the planes of constant phase 
and therefore to the group velocity ${\bf c_g}$. Viscosity damps the 
wave amplitude as
\begin{eqnarray} \label{eq:decay}
\exp\left(-\frac{\nu k^2}{|{\bf c_g}|}\xi\right)=\exp\left(-\frac{\nu k^3}{2 \Omega \sin \theta}\xi\right)
\end{eqnarray}
in the direction of the group velocity ${\bf c_g}$ at which the energy of the wave propagates~\cite{Lighthill1967,Sutherland2010} ($\xi$ is the spatial coordinate in the direction of the group velocity, see Fig.~\ref{fig:plane_wave}).

\subsection{The helical basis decomposition}

Following Smith \& Waleffe~\cite{Smith1999}, we can decompose any divergence-free velocity field ${\bf u}({\bf x},t)$ on the basis of helical modes as 
\begin{eqnarray}
{\bf u}({\bf x},t) = \sum_{\bf k} \sum_{s_{\bf k}=\pm 1} b_{s_{\bf k}}({\bf k},t) {\bf h}_{s_{\bf k}}({\bf k}) \, e^{i[{\bf k}\cdot{\bf x} - \sigma_{s_{\bf k}}({\bf k})t]} \, , \label{eq:decomp_helical}
\end{eqnarray}
where $\sigma_{s_{\bf k}}({\bf k})$ is the angular frequency of the mode with 
wavevector ${\bf k}$, amplitude $b_{s_{\bf k}}({\bf k},t)$ and polarity $s_{\bf k}=\pm 1$. The helical base vectors ${\bf 
h}_{s_{\bf k}}({\bf k})$ are defined as
\begin{eqnarray}
{\bf h}_{s_{\bf k}} = \frac{\bf k}{|{\bf k}|} \times \frac{{\bf k} \times \hat{\bf z}}{| 
	{\bf k} \times \hat{\bf z}  |} + i \, s_{\bf k} \frac{{\bf k} \times \hat{\bf z}}{| 
	{\bf k} \times \hat{\bf z}  |} \, .
\end{eqnarray}
Inserting the decomposition (\ref{eq:decomp_helical}) into the Navier-Stokes equation (\ref{eq:nsrot}) yields a set of equations for the evolution of the amplitude 
$b_{s_{\bf k}}$ of the modes $({\bf k},s_{\bf k},\sigma_{s_{\bf k}})$
\begin{eqnarray}
\left(\frac{\partial}{\partial t} +\nu k^2\right) b_{s_{\bf k}} = \frac{1}{2} \sum_{{\bf k}+{\bf p}+{\bf q}=\boldsymbol{0}} C_{{\bf k}{\bf p}{\bf q}}^{s_{\bf k} s_{\bf p} s_{\bf q}} \overline{b}_{s_{\bf p}} \, \overline{b}_{s_{\bf q}} e^{i (\sigma_{s_{\bf k}}+\sigma_{s_{\bf p}}+\sigma_{s_{\bf q}}) t} \,, \label{eq:NSdecomp}
\end{eqnarray}
due to non-linear interactions with couples of modes $({\bf p},s_{\bf 
p},\sigma_{s_{\bf p}})$ and $({\bf q},s_{\bf q},\sigma_{s_{\bf q}})$ (the overline indicates complex conjugate). In Eq.~(\ref{eq:NSdecomp}), the sum is taken over all 
wavevectors ${\bf p}$ and ${\bf q}$ such that ${\bf k}+{\bf p}+{\bf 
q}=\boldsymbol{0}$, and over the polarities $s_{\bf p}= \pm 1$, $s_{\bf q}= \pm 
1$. The triadic interaction coefficients are defined as
\begin{eqnarray}
C_{{\bf k}{\bf p}{\bf q}}^{s_{\bf k} s_{\bf p} s_{\bf q}} = \frac{s_q q - s_p p}{2} \,  [{\overline{\bf h}}_{s_p}({\bf p}) \times {\overline{\bf h}}_{s_q}({\bf q}) ] \cdot {\overline{\bf h}}_{s_k}({\bf k}) \, ,\label{eq:ckpq}
\end{eqnarray}
where $p=|{\bf p}|$ and $q=|{\bf q}|$.

\subsection{Triadic resonance of inertial waves}

In Eqs.~(\ref{eq:decomp_helical}) and (\ref{eq:NSdecomp}), if the frequency $\sigma_{s_{\bf k}}({\bf k})$ of a mode with wavevector ${\bf k}$ obeys the dispersion relation of inertial waves,
\begin{eqnarray}\label{eq:dispersion_bis}
\sigma_{s_{\bf k}}({\bf k}) =  2 \Omega \,  s_{\bf k} \, \frac{k_z}{k} \, ,
\end{eqnarray}
the resulting helical mode corresponds exactly to a plane inertial wave~\cite{Smith1999}. 
To describe the triadic resonance instability of a plane inertial 
wave, we therefore restrict the system of equations~(\ref{eq:NSdecomp}) to 
three waves defined by $({\bf k_0},s_0,\sigma_0)$, $({\bf 
k_1},s_1,\sigma_1)$ and $({\bf k_2},s_2,\sigma_2)$. These three waves have frequencies that follow 
the dispersion relation~(\ref{eq:dispersion_bis}) and that fulfill the triadic 
resonance conditions
\begin{eqnarray}
\sigma_0+\sigma_1+\sigma_2&=&0\, , \label{eq:temp_res} \\
{\bf k_0}+{\bf k_1}+{\bf k_2}&=&{\bf 0}\,. \label{eq:spat_res} 
\end{eqnarray}

The spatial resonance condition~(\ref{eq:spat_res}) was already included in 
Eq.~(\ref{eq:NSdecomp}). Then, writing the temporal resonance 
condition~(\ref{eq:temp_res}) amounts to assume that the flow is weakly 
non-linear, i.e., that the wave period $1/\sigma_i$ is much shorter than the 
non-linear time $1/(k_i b_i)$, where $b_i$ is the amplitude of wave $i$ and 
$k_i=|{\bf k_i}|$. This weak non-linearity condition is achieved when the 
Rossby number of the waves, $Ro_i=b_i k_i/(4\pi\Omega)$, is small compared to 
the normalized wave period $\sigma_i^*=\sigma_i/2\Omega$, which implies that 
non-linear processes driving the evolution of the amplitudes $b_i(t)$ are slow 
compared to the wave oscillations. Under this weak non-linearity assumption, 
the dominant contributions to the right hand side term of 
Eq.~(\ref{eq:NSdecomp}) come from waves that meet the temporal resonance 
condition such that $e^{i (\sigma_{s_{\bf 
k}}+\sigma_{s_{\bf p}}+\sigma_{s_{\bf q}}) t}=1$~\cite{Smith1999,Monsalve2020}. For non-resonant triads, the contribution of the complex exponential tends toward zero when integrated over times longer than $1/(\sigma_{s_{\bf k}}+\sigma_{s_{\bf p}}+\sigma_{s_{\bf q}})$, strongly reducing the efficiency of the energy exchanges within the triad~\cite{Galtier2003,Nazarenko2011}. Although these arguments suggest that only resonant triads are of interest, this is strictly true only at vanishing Rossby number and recent works have shown that near-resonant~\cite{LeReun2020} and even non-resonant~\cite{Brunet2020} triads can trigger instabilities of inertial waves toward 2D vertically invariant modes at finite Rossby number, these instabilities having however growth rates $Ro$ times smaller than of the triadic resonance instability.

\subsection{Instability growth rate}\label{sec:instgr}

In the following, we consider that the flow is composed of a plane inertial wave (labelled $0$), as the base flow, and two secondary plane inertial waves (labelled $1$ and $2$) that result from the instability. Without loss of generality, we choose the primary wave $0$ to be invariant in the $y$-direction, i.e., that ${\bf k_0}=(k_{x,0},0,k_{z,0}$), to have a positive angular frequency $\sigma_0$, and a negative polarity $s_0=-1$. We also consider the case (corresponding to the experiments presented later) where $k_{x,0}<0$, $k_{z,0}$ being negative as a consequence of the dispersion relation. We restrict our analysis to the early development of the instability, assuming that the primary wave amplitude 
$b_0$ remains constant and the amplitudes of the secondary waves, $b_1$ and 
$b_2$, remain small compared to $b_0$, a situation sometimes called the 
``pump-wave approximation'' \cite{Craik1978,Gururaj2020}. 
Following~(\ref{eq:NSdecomp}), the evolution of the amplitudes of the secondary 
waves is described by
\begin{eqnarray}
\frac{d b_1}{dt}=C_1 \overline{b}_0 \overline{b}_2 -\nu k_1^2 b_1 \, ,\label{eq:b1}\\
\frac{d b_2}{dt}=C_2 \overline{b}_0 \overline{b}_1 -\nu k_2^2 b_2 \, 
,\label{eq:b2}
\end{eqnarray}
where $k_1=|{\bf k_1}|$, $k_2=|{\bf k_2}|$, $C_1=C_{{\bf k_1}{\bf k_0}{\bf k_2}}^{s_{\bf k_1} s_{\bf k_0} s_{\bf k_2}}$ and $C_2=C_{{\bf k_2}{\bf k_1}{\bf k_0}}^{s_{\bf k_2} s_{\bf k_1} s_{\bf k_0}}$.
Solving this system of equations leads to an 
exponential growth (or decay) of $b_1$ and $b_2$ at a rate
\begin{equation}\label{eq:growthrate}
\gamma=\frac{-\nu(k_1^2+k_2^2)}{2}+\sqrt{\frac{\nu^2(k_1^2-k_2^2)^2}{4}+ C_1 \overline{C_2} |b_0| ^2}\, ,
\end{equation}
the product $C_1 \overline{C_2}$ being real. We first note that the instability growth rate $\gamma$ does not depend on the rotation rate of the fluid $\Omega$. Thus, $\gamma$ will depend on the primary wave Reynolds number $Re_0=b_0 2\pi/(k_0 \nu)$ and on its non-dimensional frequency $\sigma_0^*$ but not on the primary wave Rossby number $Ro_0=b_0 k_0/(4\pi \Omega)$.

In the inviscid limit where the primary wave Reynolds number tends to infinity, the expression of the instability growth rate reduces to $\gamma=|b_0| \sqrt{C_1 \overline{C_2}}$. In this situation, and in the 2D case where secondary waves propagates in the same vertical plane as the primary wave (i.e. $k_{y,1}=k_{y,2}=0$), the maximum growth rate is found for secondary wavenumbers much larger than the primary wavenumber such that ${\bf k_1} \simeq -{\bf k_2}$~\cite{Staquet2002,Koudella2006,Bordes2012}. As a consequence, the secondary waves are found at degenerated frequencies $|\sigma_1| \simeq |\sigma_2|$ equal to half the primary wave frequency $\sigma_0/2$ and the TRI is often called parametric subharmonic instability (PSI)~\cite{Staquet2002,Koudella2006,Bordes2012}.

Building on several identities pointed out in~\cite{Waleffe1992,Smith1999} (see Appendix~\ref{app:gammaRe0inf}), the inviscid growth rate can in the general 3D case be rewritten as
\begin{equation}
    \gamma=|b_0| \left[\frac{\sin^2\alpha_2}{4k_2^2}\left(s_0k_0+s_1k_1+s_2k_2\right)^2\frac{\sigma_1\sigma_2}{\sigma_0^2}\left(s_2k_2-s_1k_1\right)^2\right]^{1/2}\, ,
\end{equation}
where $\alpha_2$ is the angle opposite to the side ${\bf k_2}$ in the closed triangle formed by the resonant triads $({\bf k_0},{\bf k_1},{\bf k_2})$.

We assume in the following of this subsection~\ref{sec:instgr} that, in the inviscid 3D case, the secondary wavenumbers associated to the maximum growth rate are also much larger than the primary wavenumber such that $|\sigma_1|=|\sigma_2|=\sigma_0/2$ and $k_1\simeq k_2 \gg k_0$ (this assumption will be validated in the next subsection). Focusing on the combination of wave polarities $(s_0=-1,s_1=+1,s_2=-1)$, an asymptotic expansion of the growth rate to the first order in $k_0/k_1 \simeq k_0/k_2$ leads to a simple expression for the growth rate depending only on the angle $\alpha_2$ (see Appendix~\ref{app:gammaRe0inf})
\begin{equation}
   \frac{\gamma}{k_0\vert b_0\vert}\simeq\frac{1}{2}\sin\alpha_2\left(1-\cos\alpha_2\right)\, .
   \label{eq:gammasina}
\end{equation} 
Maximizing this expression with respect to the angle $\alpha_2$, the maximum growth rate $\gamma^{(\rm max)}/(k_0|b_0|)\simeq 0.6495$ is found for the specific angle $\alpha_2=2\pi/3$ (120$^\circ$) independently of the primary wave frequency $\sigma_0^*$.

Remarkably, this angle $\alpha_2=2\pi/3$ can only be found for 3D resonant 
triads with the secondary waves propagating in other vertical planes than the one of the 
primary wave, i.e., with a non-zero $y$-component of their wavevectors 
$k_{y,1}=-k_{y,2}$. More precisely, we show in Appendix~\ref{app:gammaRe0inf} 
that 
\begin{equation}\label{eq:kykx}
   \frac{k_{y,1}}{k_{x,1}}=\pm\left(\frac{3}{1-{\sigma^*_0}^2}\right)^{1/2} \,.
\end{equation} 
For the primary wave non-dimensional frequency that will further be considered in our experiments, $\sigma_0^*=0.84$, this corresponds to an angle
\begin{equation}
   \phi_1=\tan^{-1}\left(\frac{3}{1-{\sigma^*_0}^2}\right)^{1/2} \simeq 73^{\circ}
\end{equation} 
between the vertical plane of propagation of the primary wave and the one of secondary wave~$1$ (and secondary wave $2$ actually). For $\sigma_0^*$ ranging from $0$ to $1$, the angle $\phi_1$ ranges from $60^{\circ}$ to $90^{\circ}$. This result is remarkable: It demonstrates that, in the inviscid limit, the most unstable triad is always three dimensional with secondary waves propagating in vertical planes making an angle between $60^{\circ}$ and $90^{\circ}$ with the primary wave vertical plane of propagation.

In the following, we come back to the general 3D case with viscosity and identify numerically the maximum growth rate of the instability.

\subsection{Instability growth rate in the viscous case}

\begin{figure}
	\centerline{\includegraphics[width=13cm]{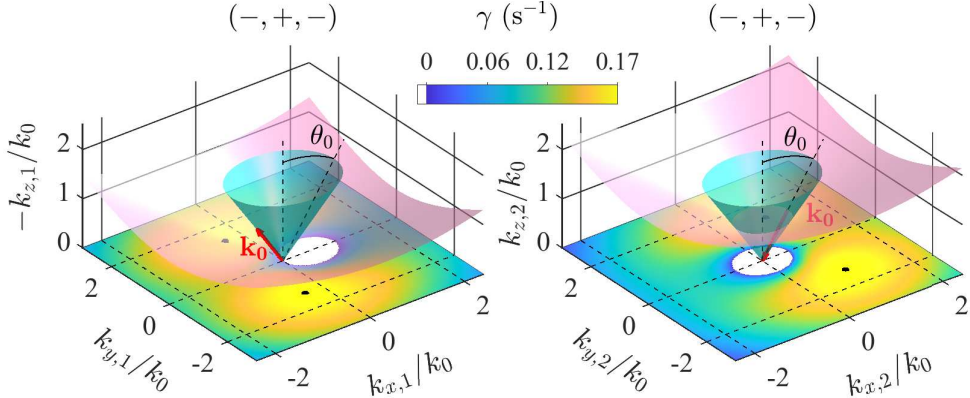}}
	\caption{Resonance surfaces of ${\bf k_1}$ (left) and ${\bf k_2}$ (right) for the combination of wave polarities $(-,+,-)$ and a primary wave defined by ($k_0=0.83$~rad/cm, $s_0=-1$, $\sigma_0^*=0.84$, $b_0=0.39$~cm/s). In the left panel, the vertical axis shows $-k_{z,1}$ for sake of clarity. The two 3D plots also show the cone of apex at ${\bf k_i}={\bf 0}$ and of semi-angle $\theta_0=\cos^{-1}(\sigma_0^*)$. This cone represents the waves at the forcing frequency according to the dispersion relation. Below each resonance surface, we also report the map of the growth rate $\gamma$ of the instability in the plane ($k_{x,i}, k_{y,i}$) in which the locations of the maximum growth rate are shown by black dots. The value $\nu=1.20\times10^{-6}$~m$^2$/s is used for the kinematic viscosity in order to match the experimental value in the next section.}\label{fig:kz1_gtheo}
\end{figure}

In Fig.~\ref{fig:kz1_gtheo}, we report in the coordinate system ($k_{x}/k_0, 
k_{y}/k_0, k_{z}/k_0$) the ``resonance surfaces'' defined by all the 
wavevectors ${\bf k_1}$ (left) and ${\bf k_2}$ (right) which are solutions of 
the triadic resonance conditions (\ref{eq:temp_res}-\ref{eq:spat_res}). These 
surfaces are computed for a primary wave defined by ($k_0=0.83$~rad/cm, 
$s_0=-1$, $\sigma_0^*=0.84$, $b_0=0.39$~cm/s), and for a combination of wave 
polarities $(s_0=-1,s_1=+1,s_2=-1)$ denoted in the following by the short-hand 
$(-,+,-)$. These resonance surfaces illustrate all the possible couples of 
secondary waves (${\bf k_1},{\bf k_2=\bf k_0-\bf k_1}$) in triadic resonance 
with the primary wave for the case $(-,+,-)$. These resonance surfaces are the 
three-dimensional extensions of the 2D classical resonance curves, reported in 
several works \cite{Smith1999,Koudella2006,Joubaud2012,Bordes2012}, which are
restricted to secondary waves invariant in the $y$-direction ($k_y=0$), as the 
primary wave. Below each resonance surface, we report the map of the 
corresponding growth rate $\gamma$ of the instability (\ref{eq:growthrate}) in 
the plane ($k_{x,i}, k_{y,i}$). The primary wave Reynolds number is $Re_0=b_0 
2\pi/(k_0 \nu) \simeq 245$ where the value $\nu=1.20\times10^{-6}$~m$^2$/s is 
used for the kinematic viscosity in order to match the experimental value in 
the next section. The rotation rate $\Omega=18$~rpm is the same as 
the one of the experiments presented later in the article. It yields a primary 
wave Rossby number of $Ro_0=b_0 k_0/(4\pi \Omega) \simeq 0.014$.

The left panel of Fig.~\ref{fig:kz1_gtheo} shows the resonance surface for the wavevector ${\bf k_1}$ (note that the vertical axis reports $-k_{z,1}$ for sake of clarity). This resonance surface (in pink) extends up to infinity, 
with any choice of the wavevector components ($k_{x,1},k_{y,1}$) having a 
resonant solution. The resonance surface uniformly lies below the cyan cone of 
apex at ${\bf k_1}={\bf 0}$ [a point also included in the resonance surface] 
and semi-angle $\theta_0=\cos^{-1}(\sigma_0^*)$ corresponding to wavevectors 
${\bf k_1}$ of waves at the forcing frequency according to the dispersion 
relation. This observation means that the resonant secondary waves $1$ always 
have a wavevector ${\bf k_1}$ which is more horizontal than the primary 
wavevector ${\bf k_0}$ (shown in the figure). According to the dispersion 
relation~(\ref{eq:dispersion}), it implies that $|\sigma_1|<|\sigma_0|$. We 
also note that a small portion of the ${\bf k_1}$ resonance surface associated 
to positive values of $k_{z,1}$ and small values of ($k_{x,1}, k_{y,1}$) is not 
shown. One can however see that this range of secondary wavevectors ${\bf k_1}$ 
is associated with a negative growth rate $\gamma$ and is therefore not to be 
further considered.

The right panel of Fig.~\ref{fig:kz1_gtheo} shows the resonance surface for the 
wavevector ${\bf k_2}={\bf k_0}-{\bf k_1}$. Although this information is 
equivalent to the one found on the left panel, the ${\bf k_2}$ resonance 
surface highlights interesting features. For instance, contrary to the case of 
${\bf k_1}$, the ${\bf k_2}$ resonance surface does not intersect the $k_z=0$ 
plane since $k_{z,2}$ is always strictly positive. In addition, a portion of 
the resonance surface at small ($k_{x,2}, k_{y,2}$) lies inside the cone of 
semi-angle $\theta_0=\cos^{-1}(\sigma_0^*)$. One can see in the map of $\gamma$ 
reported in the plane ($k_{x,2}, k_{y,2}$) that the corresponding values of 
${\bf k_2}$ are however associated with the regime of negative growth rate 
already noticed in the left panel for the wavevector ${\bf k_1}$. Again, if we 
restrict our analysis to positive growth rates $\gamma$, the secondary 
wavevectors ${\bf k_2}$ are more horizontal than the primary wavevector ${\bf 
k_0}$. In summary, the resonant secondary waves $1$ and $2$ have frequencies 
that are always smaller in absolute value than the one of the primary wave, 
$|\sigma_{1,2}|<|\sigma_0|$, i.e., the secondary wavevectors ${\bf k_1}$ and 
${\bf k_2}$ are always more horizontal than the primary wavevector~\cite{Smith1999}. Considering 
that in the present conventions frequencies can be negative, a resonant triad 
with a primary wave of positive frequency $\sigma_0$ will have negative 
frequencies for both secondary waves obeying the relation 
$|\sigma_1|+|\sigma_2|=|\sigma_0|$.

We now come to the most important conclusion of this section: the growth rate 
of the instability $\gamma$ is maximum for secondary waves which are not 
propagating in the same plane as the primary wave, i.e., for secondary waves 
that are not invariant in the $y$-direction ($k_y \neq 0$). The triadic 
resonance instability of a plane inertial wave is therefore expected to be a 
three-dimensional instability transferring energy to secondary waves with a 
wavevector component $k_y$ of the same order as $k_x$. This can be observed in 
Fig.~\ref{fig:kz1_gtheo}, where the two locations of the maximum growth rate in 
the ($k_{x,i}, k_{y,i}$)-planes are indicated by black dots. There are actually 
two couples of secondary waves that maximize the growth rate with symmetric 
wavevectors ${\bf k_i}$ with respect to the primary wave vertical plane 
$k_{y,i}=0$. In the sequel, we will show that this three-dimensionality is in 
agreement with the experiments.

\begin{figure}
	\centerline{\includegraphics[width=14cm]{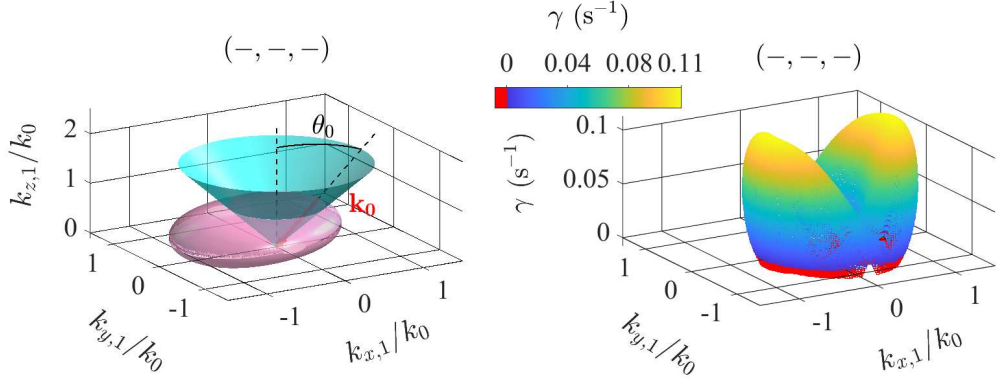}}
	\caption{Resonance surface of ${\bf k_1}$ (left, in pink) and 3D view of the instability growth rate $\gamma$ (right) for the combination of polarities $(-,-,-)$. In the left panel, we also show the cone of apex at ${\bf k_1}={\bf 0}$ and of semi-angle $\theta_0=\cos^{-1}(\sigma_0^*)$. The surfaces are computed for a primary wave defined by ($k_0=0.83$~rad/cm, $s_0=-1$, $\sigma_0^*=0.84$, $b_0=0.39$~cm/s) in a rotating fluid of kinematic viscosity $\nu=1.20\times10^{-6}$~m$^2$/s.}\label{fig:kz1_gtheo_2}
\end{figure}

For the sake of completeness, we also consider the combinations of polarities 
$(-,-,+)$ and $(-,-,-)$. The first combination, $(-,-,+)$ is actually the same 
case as $(-,+,-)$ where the roles of the waves $1$ and $2$ have been exchanged. 
For the polarities combination $(-,-,-)$, Fig.~\ref{fig:kz1_gtheo_2} shows the 
resonance surface for ${\bf k_1}$ on the left and a 3D view of the 
corresponding growth rate $\gamma$ as a function of ($k_{x,1}, k_{y,1}$) on the 
right. This representation is necessary since the resonance surface is a closed 
surface with values of $k_1$ of the order of $k_0$. This implies that for a 
given couple of wavevector components ($k_{x,1}, k_{y,1}$) there are either two 
resonant solutions for $k_{z,1}$ at small ($k_{x,1}, k_{y,1}$) or no solution 
at large ($k_{x,1}, k_{y,1}$). Then, $\gamma$ takes also two values when 
$k_{z,1}$ does. As for the $(-,+,-)$ instability, when the instability growth 
rate $\gamma$ is positive, the secondary waves are subharmonic with 
$|\sigma_1|$ and $|\sigma_2|$ smaller than $|\sigma_0|$. On the contrary, the 
maximum growth rate of the instability is this time found for secondary waves 
with $k_y=0$, invariant along the $y$-direction. At the considered 
non-dimensional frequency $\sigma_0^*=0.84$, the $(-,-,-)$ instability is 
therefore 2D to the first order, i.e., if one considers only its maximum growth 
rate.

\begin{figure}
	\centerline{\includegraphics[width=10cm]{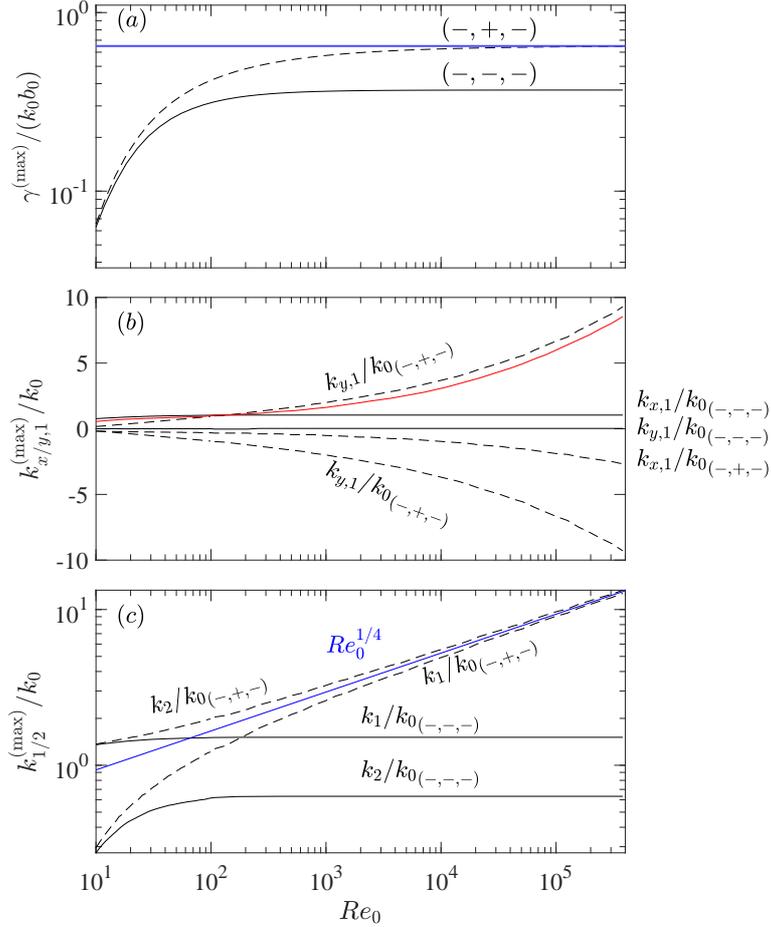}}
	\caption{(a) Maximum normalized instability growth rate $\gamma^{\rm 
	(max)}/(k_0 b_0)$ for polarities combinations $(-,+,-)$ and $(-,-,-)$ as a 
	function of the primary wave Reynolds number $Re_0=b_0 2\pi/(\nu k_0)$ for 
	$\sigma_0^*=0.84$. The horizontal blue straight line reports the value 
	$\gamma^{\rm (max)}/(k_0 b_0)$ predicted analytically in the inviscid limit 
	in the previous section~\ref{sec:instgr} (see Eq.~\ref{eq:gammasina}). (b) 
	Wavevector components $k_{x,1}^{\rm (max)}$ and $k_{y,1}^{\rm (max)}$ 
	corresponding to the maximum growth rate for each mode as a function of 
	$Re_0$ (again for $\sigma_0^*=0.84$). The red curve shows the wavenumber 
	$\sqrt{3/(1-{\sigma_0^*}^2)}\,|k_{x,1}^{\rm (max)}|$ which is predicted in 
	the inviscid limit (see section~\ref{sec:instgr}) to match the wavevector 
	component $k_{y,1}^{\rm (max)}$ for the instability mode $(-,+,-)$. (c) 
	Corresponding wavenumbers. The blue straight line shows a power law 
	$Re_0^{1/4}$.}\label{fig:gamma_Re}
\end{figure}

To conclude this section, we report in Fig.~\ref{fig:gamma_Re}(a) the evolution 
of the maximum instability growth rate $\gamma^{\rm (max)}$ for combinations 
$(-,+,-)$ and $(-,-,-)$ as a function of the primary wave Reynolds number 
$Re_0=b_0 2\pi/(\nu k_0)$ for $\sigma_0^*=0.84$. This maximum growth rate is actually shown 
normalized by the non-linear frequency of the primary wave as $\gamma^{\rm 
(max)}/(k_0 b_0)$. The normalized maximum growth rate naturally grows with 
$Re_0$ for both modes starting from vanishing and asymptotically equal values 
at small $Re_0$. At large $Re_0$, the normalized growth rate $\gamma^{\rm 
(max)}/(k_0b_0)$ tends toward order~$1$ asymptotic values, the growth rate for 
the $(-,+,-)$ mode being typically twice larger than for the $(-,-,-)$ mode. 
Moreover, in the $(-,+,-)$ case, the maximum normalized growth rate tends 
toward the inviscid value $\gamma^{(\rm max)}/(k_0 b_0)\simeq 0.6495$ 
predicted analytically  in the previous section~\ref{sec:instgr} (see 
Eq.~\ref{eq:gammasina}). In Fig.~\ref{fig:gamma_Re}(b), we show the wavevector 
components $k_{x,1}^{\rm (max)}$ and $k_{y,1}^{\rm (max)}$ corresponding to the 
maximum growth rate for each mode and as a function of $Re_0$ (again for 
$\sigma_0^*=0.84$). For the instability mode $(-,+,-)$, $k_{x,1}^{\rm (max)}$ 
is negative and slowly grows in absolute value from $\sim k_0/10$ at 
$Re_0\simeq 10$ up to $\sim 2.7\, k_0$ at $Re_0\simeq 4\times 10^5$ whereas 
$k_{y,1}^{\rm (max)}$ can take two opposite values that grow in absolute value 
from $\sim k_0/10$ at $Re_0\simeq 10$ to $\sim 9.3\, k_0$ at $Re_0\simeq 
4\times 10^5$, in agreement with the symmetry found in 
Fig.~\ref{fig:kz1_gtheo}. We also report in Fig.~\ref{fig:gamma_Re}(b) as a red 
curve the wavenumber $|k_{x,1}^{\rm (max)}|\,\sqrt{3/(1-{\sigma_0^*}^2)}$ which 
is predicted in the inviscid limit (see section~\ref{sec:instgr}) to match the 
wavevector component $k_{y,1}^{\rm (max)}$. One can observe that the value of 
$k_{y,1}^{\rm (max)}$ numerically obtained from the full set of viscous 
equations
is actually already close to its 
inviscid prediction at moderate $Re_0$. The behavior of the mode $(-,-,-)$ is 
very different. The maximum-growth-rate 
instability is 2D, with $k_{y,1}^{\rm (max)}=0$ for all $Re_0$, and 
$k_{x,1}^{\rm (max)}$ positive, of order $k_0$ and slowly increasing with 
$Re_0$. Figure~\ref{fig:gamma_Re}(c) finally shows that for the $(-,+,-)$ mode 
the norm of the subharmonic wavenumbers, $k_1$ and $k_2$, continuously grows 
from values of the order of $\sim 0.3\, k_0$ and $\sim 1.4\, k_0$ at 
$Re_0\simeq 10$, respectively, up to values of the order of $\sim 13 \, 
k_0$ at $Re_0\simeq 4\times 10^5$. At large $Re_0$, $k_1/k_0$ and $k_2/k_0$ 
increase following power laws $Re_0^{1/4}$. In parallel, for 
the $(-,-,-)$ mode, $k_1$ and $k_2$ slowly grow over the considered range of 
Reynolds number while remaining in the range between $0.3\, k_0$ and 
$1.5\, k_0$.

The previous theoretical developments demonstrate the 3D character of the TRI 
of a plane inertial wave. In the sequel, we explore this question from an
experimental point of view before we finally compare the two approaches 
quantitatively.

\section{Experimental setup}\label{sec:setup}

\begin{figure}
    \centerline{\includegraphics[width=10cm]{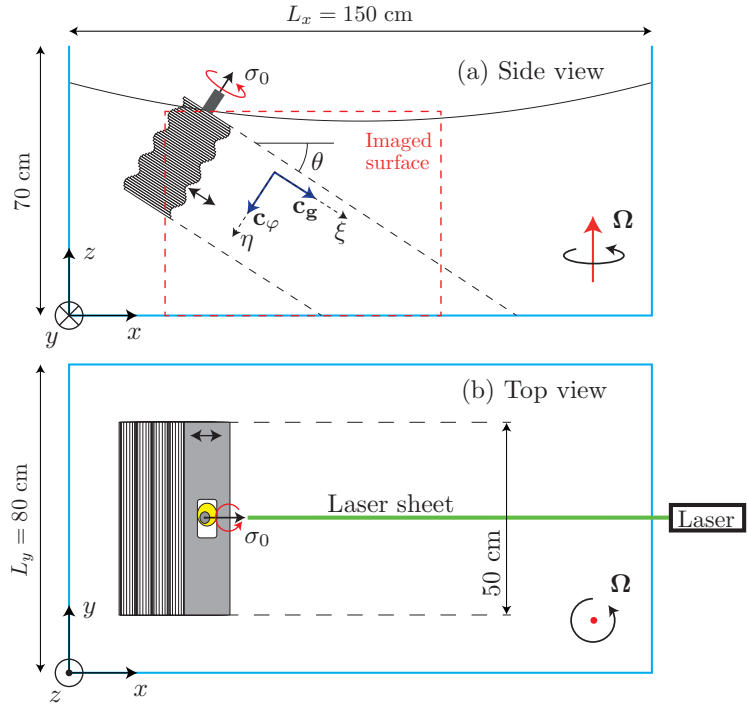}}
    \caption{Sketch of the experimental setup seen from the side (a) and from 
    the top (b).}\label{fig:cavite}
\end{figure}

The flow is generated in a parallelepipedic glass tank of $L_x \times L_y=150 
\times 80$~cm$^2$ rectangular base and $70$~cm height filled with $55$~cm of 
water as sketched in Fig.~\ref{fig:cavite}. A plane inertial wave is forced in 
the tank by an immerged wave maker which has already been implemented in 
several studies of internal gravity waves in stratified 
fluids~\cite{Joubaud2012, Scolan2013, Bourget2013, Brouzet2017} and in a 
previous study of the instability of a plane inertial wave~\cite{Bordes2012}. A 
major difference with these previous experiments has nevertheless been 
introduced : the wake maker produces here a plane wave with a large spatial 
extension in the horizontal direction $y$ (in which the forced wave is supposed 
to be invariant) normal to the wave propagation plane ($x,z$). More precisely, the wave maker 
extent in the $y$-direction is of $50$~cm whereas it was of $14$~cm in 
the previous studies, these lengths being to be compared to the forced 
wavelength of $7.6$~cm.

The wave maker is composed of a stack of $48$~plates which are $6.33$~mm thick 
and $50$~cm wide (see Fig.~\ref{fig:cavite}). The plates are fitted with a 
rectangular hole at their center through which a camshaft is inserted, each cam 
being a circular plate adjusted to the hole with a rotation point shifted from 
its center by an eccentricity $A$ (scotch yoke mechanism). A constant angular 
shift of $30$~degrees is introduced between adjacent cams leading the surface 
drawn by the plate edges to approximate a sinusoidal shape of wavelength 
$\lambda_f=7.6$~cm. The profile contains four wavelengths such that the 
produced wave beam will have a $4\lambda_f = 30.4$~cm width. A brushless motor 
coupled to a reducer is driving the camshaft in a constant rotation at an 
angular frequency $\sigma_0$ such that each plate is finally subject to an 
oscillating linear translation motion in the direction normal to its width and 
to the camshaft axis (the plates are guided laterally). The wave maker surface 
eventually describes a sinus profile
\begin{equation}\label{eq:wavemaker}
    \xi_{\rm wm}(\eta,t)=A\sin(\sigma_0 t - k_0 \eta)\, ,
\end{equation}
with a phase propagating downward, parallel to the camshaft axis. In Eq.~(\ref{eq:wavemaker}), $k_0=2\pi/\lambda_f$ is the wavenumber, $\eta$ is the coordinate along the phase propagation direction and $\xi$ the coordinate along the energy propagation direction (see Fig.~\ref{fig:cavite}).

The whole system is mounted on a $2$-m diameter platform rotating at a rate 
$\Omega=18$~rpm. The angular frequency of the wave maker is set to 
$\sigma_0=0.84 \times 2 \Omega \simeq 3.17$~rad/s. Following the inertial wave 
dispersion relation, the wave maker is tilted at an angle 
$\theta_0=\cos^{-1}(\sigma_0/2\Omega) \simeq 32.9^{\circ}$ with its deforming 
surface pointing downwards. With this tilt, the motion of the wave maker 
surface matches the velocity boundary condition of a plane inertial wave at 
frequency $\sigma_0$ and propagating downwards (polarity $s_0=-1$). More 
precisely, the wave maker drives inertially the velocity component of the plane 
wave along its energy propagation direction (axis $\xi$ in 
Fig.~\ref{fig:cavite}) without however forcing the velocity component along the 
wave invariance direction $y$. Given the location of the wave maker in the tank 
(Fig.~\ref{fig:cavite}), the forced wave will propagate over a distance of 
about $60$~cm before the reflection on the bottom of the tank takes place. In 
our study, we use cams with eccentricity $A$ equal to either $1$, $1.5$ or 
$2$~mm, leading to forcing Reynolds numbers $Re_f=A\sigma_0\lambda_f/\nu$ in 
the range $230 \leq Re_f \leq 420$ and forcing Rossby numbers 
$Ro_f=A\sigma_0/(2\Omega \lambda_f)$ in the range $0.011 \leq Ro_f \leq 0.022$.

The two components $(u_x,u_z)$ of the velocity field are measured in the 
vertical plane $y=y_0=L_y/2$ using a particle image velocimetry (PIV) system 
mounted in the rotating frame ($y=0$ is the front side of the tank). The water 
is seeded with $10$~$\mu$m tracer particles and illuminated by a laser sheet 
generated by a corotating 140~mJ Nd:YAG pulsed laser. For each experiment, 
$7\,920$~images of particles are acquired using a $2\,360 \times 1\,776$~pixels 
camera at a frequency of $24$~images per wave maker period $T=2\pi/\sigma_0$. The acquisition, 
which is started $30$ forcing periods before the start of the wave maker, 
covers $330$~periods in total. The imaged region has a surface of $71 
\times 53$~cm$^2$ (see the
dashed rectangle in Fig.~\ref{fig:cavite}). PIV cross-correlation is finally 
performed between successive images using $32 \times 32$~pixels interrogation 
windows with a $50\%$ overlap and provides velocity fields with a spatial 
resolution of $4.8$~mm. The rotation of the platform is always started at least 
$30$~min before the start of the wave maker in order for the spin-up of the 
fluid to be completed.

\section{Experimental results}\label{sec:results}

\subsection{Subharmonic instability}\label{sec:subh}

In order to explore the temporal content of the flow produced by the wave maker, we compute the temporal power spectral density of the measured velocity field as
\begin{equation}
    E(\sigma,t,\Delta T)=\frac{4\pi}{\Delta T}\langle \vert 
    \widetilde{\mathbf{u}}(x,z,\sigma,t,\Delta T)\vert^2\rangle
    \label{eq:tpsd}
\end{equation}
where the components of $\widetilde{\mathbf{u}}$ are given by
\begin{equation}\label{eq:fft}
    \widetilde{u}_j(x,z,\sigma,t,\Delta 
    T)=\frac{1}{2\pi}\int_{t-\Delta T/2}^{t+\Delta T/2} 
    u_j(x,y_0,z,t')e^{-i\sigma t'}dt'\, ,
\end{equation}
the temporal Fourier transform of the velocity component $u_j(x,y_0,z,t)$ with 
$j=(x,z)$ and the angular brackets denote the spatial average over the 
measurement area in the plane $y_0=L_y/2$.

\begin{figure}
	\centerline{\includegraphics[width=11cm]{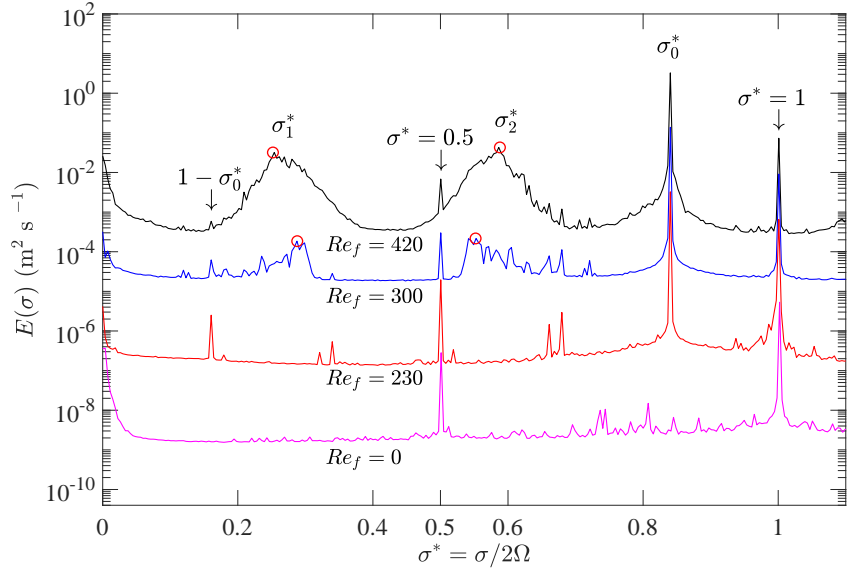}}
	\caption{Temporal power spectral density $E(\sigma,t=150\,T,\Delta
		T=300\,T)$ as a function of the normalized frequency 
		$\sigma^*=\sigma/2\Omega$ for experiments at 
		$\sigma_0^*=0.84$ and $\Omega=18~$rpm for a forcing wavelength 
		$\lambda_f=7.6$~cm and four forcing Reynolds numbers, $Re_f=0$ (pink 	
		curve), $Re_f=230$ (red), $300$ (blue) and $420$ (black). The spectrum 
			at  $Re_f=0$ corresponds to an experiment with the wave generator 
			off. A 
		vertical shift by a factor of 
		$20$ has been introduced between successive spectra.}\label{fig:spkfreq}
\end{figure}

In Fig.~\ref{fig:spkfreq}, we report the temporal spectra $E(\sigma,t,\Delta
T)$ as a function of the normalized frequency $\sigma^*=\sigma/2\Omega$ for 
the three experiments at forcing amplitudes $A=1, 1.5$ and $2$~mm. The 
Fourier transform~(\ref{eq:fft}) is computed over the whole experimental duration from the start of the wave maker ($\Delta T=300\,T$, $t=\Delta T/2$, 
$T=2\pi/\sigma_0$ being the period of the forcing). As a reference, we 
also report a spectrum measured with the wave generator off ($A=0$ and 
$Re_f=0$). All the 
spectra (with the wave generator on) exhibit an energetically 
dominant peak at the driving frequency $\sigma_0^*=0.84$. The spectrum at the 
lowest (non-zero) forcing amplitude (Reynolds number $Re_f=230$) corresponds to a flow in 
the linear regime, below the onset of the triadic resonance instability. 
Nevertheless, secondary peaks at frequencies $\sigma=\Omega$ ($\sigma^*=0.5$), $\sigma=2\Omega$ ($\sigma^*=1$)
and $\sigma=2\Omega-\sigma_0$ ($\sigma^*=0.16$) are observed. One can note that 
the energy 
peaks at $\sigma=\Omega$ ($\sigma^*=0.5$) and $\sigma=2\Omega$ ($\sigma^*=1$) are already present, with the 
same amplitude, in the spectrum without forcing ($Re_f=0$). The peak at 
$\sigma=\Omega$ ($\sigma^*=0.5$) has actually been shown to correspond mainly to a flow created 
by the rotating platform's precession induced by the Earth's rotation (see 
\cite{Boisson2012,Triana2012}). 
Nevertheless, we cannot exclude that part of the energy in this peak is 
related to mechanical perturbations of the system rotation at the frequency
$\sigma=\Omega$ inducing inertial waves in the flow.
The peak at $\sigma^*=1$ is the result of a perturbation of the 
platform rotation inducing waves at $\sigma=2\Omega$ in the water tank. The 
peak at $\sigma=2\Omega-\sigma_0$ ($\sigma^*=0.16$) can be interpreted as the result of 
the interaction between the mode at $\sigma=2\Omega$ and the forcing.
We also observe a peak at $\sigma^*=0$ with a tail extending up to 
$\sigma^*\simeq 0.05$. This peak, already present with a similar 
amplitude in the spectrum without forcing, has already been discussed in 
Refs.~\cite{Bordes2012,Brunet2019} and is due to the presence of thermal 
convection columns drifting horizontally in the water tank.
Finally, other even weakly energetic peaks are present in the spectrum at 
$Re_f=230$ corresponding to direct combinations (sums and 
differences) of the frequencies of the leading energetic modes at 
$\sigma=\sigma_0$, $\sigma=\Omega$, $\sigma=2\Omega$. As one can see, these 
modes are progressively drowned in 
the spectral noise for the experiments conducted at larger forcing amplitudes.

When increasing the forcing Reynolds number to $Re_f=300$,
two spectral bumps at subharmonic 
frequencies emerge. These bumps are almost perfectly symmetric with respect to 
half the forcing frequency $\sigma_0/2$, which indicates that the frequencies 
associated with these two bumps are in triadic resonance with the primary wave 
frequency $\sigma_0$. This observation is the classical signature of the 
triadic resonance instability of an inertial wave, and it has been widely 
reported in 
experimental~\cite{Bordes2012,LeReun2019,Brunet2019,Brunet2020,Monsalve2020} 
and numerical works~\cite{Jouve2014,LeReun2020}. As the forcing Reynolds number 
increases to $420$, the subharmonic bumps are spreading in frequency in 
agreement with previous experimental works~\cite{Brunet2019,Brunet2020}. This 
latter feature is at odds with the subharmonic peaks observed in numerical 
simulations of a 2D inertial wave attractor by Jouve and 
Ogilvie~\cite{Jouve2014} where the flow is strictly invariant in the transverse 
horizontal direction $y$. A possible explanation is that the large frequency 
width observed here for the subharmonic bumps produced by the TRI is a 
consequence of the three-dimensionality of the TRI allowed in the experiments 
but forbidden in the 2D simulations. Nevertheless, one cannot exclude that part 
of the spreading of the TRI subharmonic bumps observed here at $Re_f=420$ is 
the consequence of the emergence of secondary triadic resonant 
interactions in the flow, which could stand as the premises of a transition 
toward an inertial wave turbulence following the scenario reported 
in~\cite{Monsalve2020}.

To further explore the characteristics of the secondary waves produced by the 
triadic resonance instability, we select, for each couple of subharmonic bumps, 
the angular frequency $\sigma_1^*$ associated to the maximum spectral density 
of the lowest-frequency bump in the spectra of Fig.~\ref{fig:spkfreq}. We 
report these values in Table~\ref{tab:freqb} and highlight them in Fig.~\ref{fig:spkfreq}. Then, we select the frequency 
$\sigma_2^*=\sigma_0^*-\sigma_1^*$ in triadic resonance with $\sigma_1^*$. For 
each couple of bumps, we can see in Fig.~\ref{fig:spkfreq} that the computed 
value $\sigma_2^*$ is very close to the frequency associated to the maximum of 
the second spectral bump ($\sigma_1^*$ and $\sigma_2^*$ are shown in 
Fig.~\ref{fig:spkfreq} by red circles).

\begin{table}
	\begin{tabular}{|c | c c c|}
		\hline
		& \hspace{0.5cm}Experiment \#1\hspace{0.5cm} & 
		\hspace{0.5cm}Experiment \#2\hspace{0.5cm} & 
		\hspace{0.5cm}Experiment \#3\hspace{0.5cm} \\			
		\hline
		$Re_f$ & 230 & 300 & 420\\
		$Ro_f$ & 0.011 & 0.017 & 0.022\\
		$A$~(mm) & 1  & 1.5  & 2.0 \\
		$\nu$~(m$^2/$s)  & $1.05\times10^{-6}$  & $1.20\times10^{-6}$ & 
		$1.15\times10^{-6}$ \\
		$\sigma_1^*$  &  no TRI & 0.285  & 0.252\\
		$\sigma_2^*=\sigma_0^*-\sigma_1^*$  & no TRI & 0.555 & 0.588\\
		$k_{x,1} $ (rad/cm) &  no TRI & $0.640\pm0.104$  & 
		$0.625\pm0.110$\\	
		$k_{y,1} $ (rad/cm) &  no TRI & $1.314 \pm 0.313$  & $0.794 \pm 
		0.125$\\
		$k_{z,1}$ (rad/cm) &  no TRI &  $0.435\pm0.097$ & 
		$0.263\pm0.043$\\
		$k_{x,2}$ (rad/cm)&	no TRI & $-1.124\pm0.145$ &	$-1.057\pm 0.120$\\
		$k_{y,2} $ (rad/cm) &  no TRI & $-1.188 \pm 0.125$ & $-0.729 \pm 
		0.070$\\
		$k_{z,2}$ (rad/cm)& no TRI & $-1.091\pm0.127$ & $-0.934\pm0.058$ \\
		\hline
	\end{tabular}
	\caption{Table reporting for each experiment the normalized frequency 
		$\sigma_1^*$ associated to the maximum of the 
		bump at the lowest frequency in each subharmonic couple and the associated 
		resonant frequency $\sigma_2^*=\sigma_0^*-\sigma_1^*$ 
		($\sigma_0^*=\sigma_0/2\Omega=0.84$). These frequencies are highlighted 
		by red circles in Fig.~\ref{fig:spkfreq}. The table also report 
		for each subharmonic mode the values computed in 
		section~\ref{sec:wavevec} of the wavevector components with the 
		corresponding error bars. The 
		forcing Reynolds and Rossby 
		numbers are defined as $Re_f=A\sigma_0\lambda_f/\nu$ and 
		$Ro_f=A\sigma_0/(2\Omega\lambda_f)$, respectively. $\nu$ is the 
		experimental kinematic viscosity of water (derived from temperature 
		measurements), $\Omega=18$~rpm is the global rotation rate, $\sigma_0$ the 
		forcing frequency, $A$ the forcing amplitude, and $\lambda_f=7.6$~cm the 
		forcing wavelength.}\label{tab:freqb}
\end{table}

In the following, we use a Hilbert filtering procedure to extract the velocity field associated with the modes at frequencies $\sigma_1$ and $\sigma_2$. This procedure consists in a band-pass Fourier filter of the velocity field at the frequency of interest $\sigma_i$ (with a band width equal to the spectral resolution) in conjunction with a filtering in the wavevector space retaining only the energy present in one (judiciously chosen) quadrant of the wavevector space ($k_x,k_z$). Compared to a simple temporal Fourier filtering, this procedure allows us to remove other wave beams at the frequency of interest coming from reflections on the water tank boundaries, which have wavevectors components in the other three quadrants of the wavevector space. Moreover, the temporal filtering retains only the energy at the selected frequency $\sigma_i>0$ without including the corresponding negative frequency. This procedure leads to a complex velocity field whose real part is the physical field (once multiplied by $2$ to compensate the discarded negative frequency and enforce energy conservation) and whose argument $\varphi({\bf x})$ is the phase field of the wave of interest at the frequency $\sigma_i$. Details on this filtering procedure can be found in Refs.~\cite{Bordes2012,Brunet2019,Mercier2008}.

\begin{figure}
\centerline{\includegraphics[width=14cm]{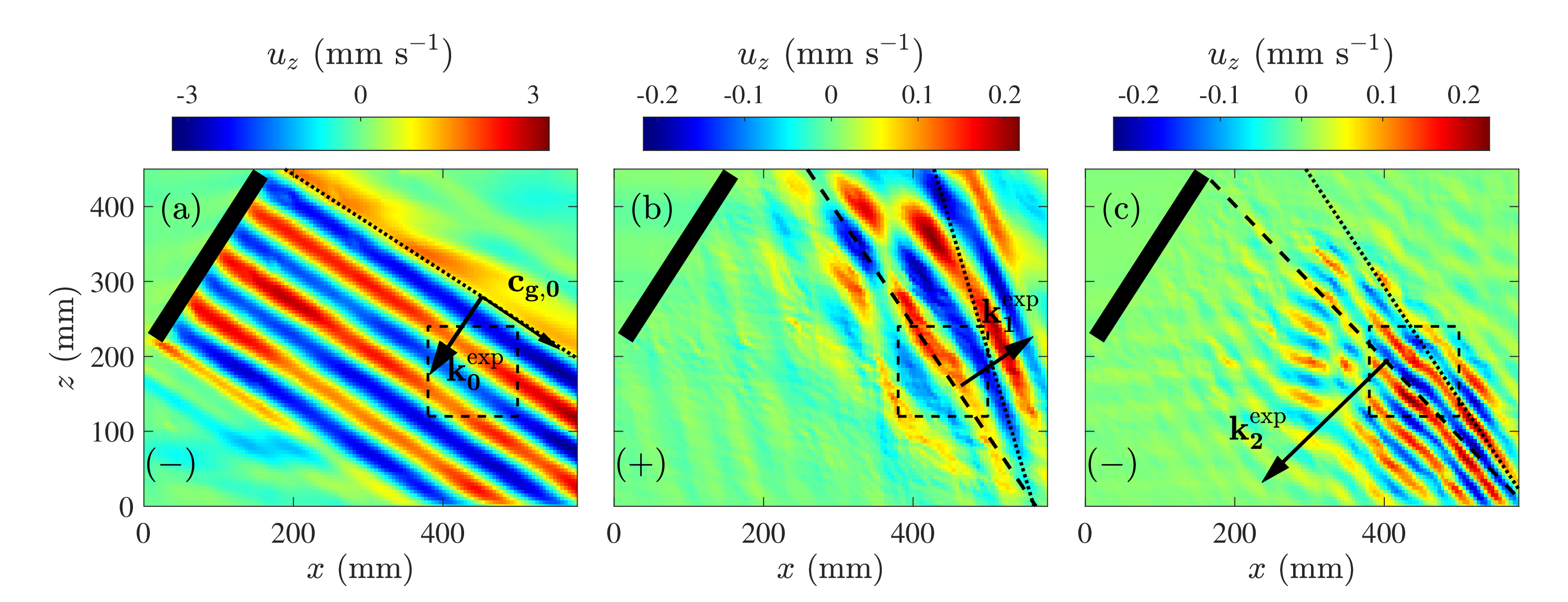}}
\caption{Snapshots of the vertical velocity component computed with the Hilbert 
filtering procedure for the experiment at $Re_f=300$ for frequencies 
$\sigma_0^*$, $\sigma_1^*$ and $\sigma_2^*$, respectively. (a) Temporal filter 
at $\sigma_0^*=0.84$ and spatial filter keeping the wavevector quadrant 
($k_x<0, k_z<0$); (b) $\sigma_1^*=0.285$ and ($k_x>0, k_z>0$); (c) 
$\sigma_2^*=0.555$ and ($k_x<0, k_z<0$). In each panel, the dotted line 
indicates the theoretical tilt angle $\theta_i=\cos^{-1}(\sigma_i^*)$ with the 
horizontal predicted for the constant phase planes for a wave at the considered 
frequency $\sigma_i^*$ and propagating in the vertical measurement plane 
($x,z$), i.e. with $k_y=0$. The black rectangle (more precisely its underside) 
indicates the mean position of the wave generator surface. In (b) and (c), 
the dashed line indicates the direction normal to the projection in the 
measurement plane of the wavevectors computed in Sec.~\ref{sec:wavevec} for the 
modes at $\sigma_1^*$ and $\sigma_2^*$ in the instability region (dashed 
rectangle).}\label{fig:vy_H_lb7p6_A1p5mm_O18rpm}
\end{figure}

For the experiment at $Re_f=300$, the panels of 
Fig.~\ref{fig:vy_H_lb7p6_A1p5mm_O18rpm} show snapshots of the vertical velocity 
field computed via the Hilbert filtering procedure for the three 
frequencies $\sigma_0^*$, $\sigma_1^*$ and $\sigma_2^*$. In panel (a), we 
observe a plane primary wave having a wavebeam width of four wavelengths, as 
expected from the wavemaker geometry shown in Fig.~\ref{fig:cavite}. The 
amplitude of the velocity oscillations $u_\xi$ of the primary wave measured 
experimentally along the energy propagation direction ${\bf c_{g,0}}$ is 
$4.0\pm 0.3$~mm/s (Fig.~\ref{fig:vy_H_lb7p6_A1p5mm_O18rpm}(a) reports the 
velocity component $u_z$ which is equal to $(1-{\sigma_0^*}^2)^{1/2}\, 
u_\xi\simeq 0.54\, u_\xi$ for an in-plane wave at frequency $\sigma_0^*=0.84$). 
This value is consistent with the forcing velocity amplitude $A_f\sigma_0 
\simeq 
4.75$~mm/s. Figures~\ref{fig:vy_H_lb7p6_A1p5mm_O18rpm} (b) and (c) show 
snapshots of the vertical velocity obtained from Hilbert filtering at the 
frequencies $\sigma_1^*$ and $\sigma_2^*$ (see Table~\ref{tab:freqb}), 
respectively. In the measurement plane, the wavelengths of the 
secondary waves are of the same order as the primary wavelength. Besides, 
the secondary waves velocity oscillation amplitude is of the 
order of $0.5$~mm/s, one order of magnitude smaller than the amplitude of the 
primary wave. The characteristic Rossby number of the secondary waves is 
therefore of the order of $10^{-3}$.

In Figs.~\ref{fig:vy_H_lb7p6_A1p5mm_O18rpm}(b) and (c), we note that the 
triadic resonance instability emerges only after the primary wave has traveled 
a distance between $20$ and $30$~cm from the wavemaker. The region, where the 
instability develops, is identified by a dashed rectangle in the fields of 
Fig.~\ref{fig:vy_H_lb7p6_A1p5mm_O18rpm}. The reason why the TRI does not 
occur closer to the wavemaker remains an open question. After analyzing the 
direction of propagation of the phase for each of the subharmonic waves, we 
conclude that 
the wave at $\sigma_1$ has a $s=+1$ polarity (upward phase propagation), 
whereas the wave at $\sigma_2$ has a $s=-1$ polarity (downward phase 
propagation): the instability observed here has a polarity combination 
$(-,+,-)$, in line with the theory presented in Sec.~\ref{sec:theory} which 
predicts that this polarity combination is associated with the maximum 
instability growth rate. Consistent with the directions of their respective 
group velocities (upward for $s=+1$ waves and downward for $s=-1$ waves), the 
wave at $\sigma_1$ spreads out upwards with respect to the instability region 
(the dashed rectangle), whereas the wave at $\sigma_2$ spreads out downwards 
with respect the instability region (see Fig.~\ref{fig:plane_wave}).

Remarkably, we observe in Fig.~\ref{fig:vy_H_lb7p6_A1p5mm_O18rpm} that the 
apparent planes of constant phase of the subharmonic waves at frequencies $\sigma_1^*$ 
and $\sigma_2^*$ (dashed line) are more horizontal than the theoretical tilt angle 
$\theta_i=\cos^{-1}(\sigma_i^*)=\cos^{-1}(k_{z,i}/k_i)$ (dotted line) expected for waves at 
the considered frequencies and propagating in the plane $(x,z)$ as the primary 
wave, i.e., invariant in the $y$-direction. This observation is an evidence 
that the subharmonic waves have a non-zero wavevector component along the 
$y$-direction, and therefore, that they are propagating out of the primary wave 
(and measurement) plane. This out-of-plane propagation explains that the 
apparent tilt angle $\cos^{-1}(k_{z,i}/(k_{x,i}^2+k_{z,i}^2)^{1/2})$ of the 
constant phase planes observed in the measurement plane is lower than the 
actual angle of the out-of-plane waves 
$\cos^{-1}(k_{z,i}/(k_{x,i}^2+k_{y,i}^2+k_{z,i}^2)^{1/2})$. The experimental 
triadic resonance instability is three-dimensional, in agreement with the 
theoretical prediction of Sec.~\ref{sec:theory}.

At this point, it is important to highlight that because we only measure 
the cut of the velocity field in a vertical plane ($y=y_0$), we are unable to 
directly measure the component of the subharmonic modes wavevectors 
along the horizontal direction $y$ normal to the measurement plane. As a 
consequence, we are not able to demonstrate that the subharmonic modes 
verify the dispersion relation of inertial waves (which involves measuring 
$k_y$ for waves propagating out of the measurement plane). Nevertheless, given 
the low Rossby number of the subharmonic modes ($\sim 10^{-3}$) and of the 
primary wave ($\sim 10^{-2}$), it is reasonable to assume that the 
subharmonic modes are indeed inertial waves. Thus, instead of showing that the 
subharmonic 
modes verify the dispersion relation, we will use the 
dispersion relation to compute the out-of-plane component of their 
wavevectors. In the next section~\ref{sec:wavevec}, we compute the wavevectors 
of the subharmonic waves at frequencies $\sigma_1^*$ and $\sigma_2^*$ and 
compare them with the theoretical predictions for the 3D TRI described in 
section~\ref{sec:theory}. The excellent agreement that we find between the 
theory and the experiments strongly supports \textit{a posteriori} the validity 
of the 
assumption that the subharmonic modes are inertial waves.

\begin{figure}
	\centerline{\includegraphics[width=12cm]{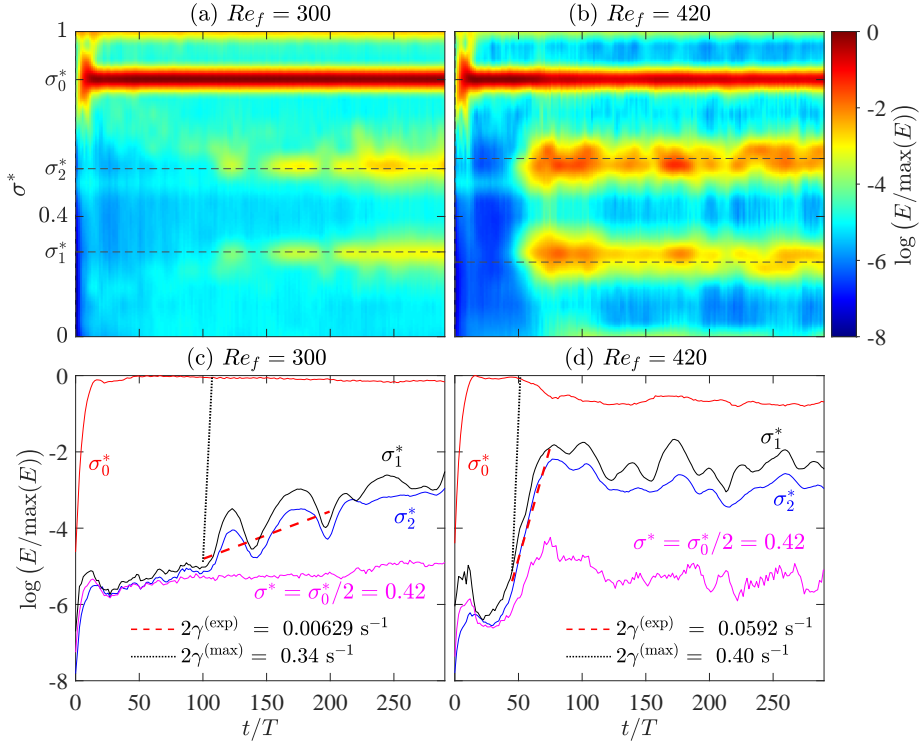}}
\caption{(a-b) Natural logarithm of the temporal energy spectra $E(\sigma,t,\Delta T)$ 
normalized by its maximum as a function of time $t$ and of the normalized 
frequency $\sigma^*=\sigma/2\Omega$ for the experiments at $Re_f=300$ (a) 
and	$Re_f=420$ (b), computed using a sliding time window of $\Delta T= 
15\,T$. (c-d) Corresponding time evolution of the temporal energy spectra 
$E(\sigma,t,\Delta T)$ normalized by its maximum for four specific frequencies, 
$\sigma_0^*=0.84$, $\sigma_1^*$, $\sigma_2^*$ (reported in 
Table~\ref{tab:freqb}) and $\sigma^*=\sigma_0^*/2$, for the experiments at 
$Re_f=300$ (c) and $Re_f=420$ (d).}\label{fig:timefreq}
\end{figure}

Before, it is important to uncover the time evolution of the flow 
from the start of the 
forcing. For this study, we focus on the region where the instability takes 
place (dashed rectangle in Fig.~\ref{fig:vy_H_lb7p6_A1p5mm_O18rpm}). We report 
in the top panels of Fig.~\ref{fig:timefreq} the natural logarithm of the 
temporal 
energy spectrum $E(\sigma,t,\Delta T)$ normalized by its maximum as a function 
of time $t$ and of the 
normalized frequency $\sigma^*=\sigma/2\Omega$ for the 
experiments at $Re_f=300$ and $Re_f=420$. These time-frequency spectra are 
computed using a short sliding time window of $\Delta T=15\,T$ in order to 
preserve the time resolution as much as possible while accessing a reasonably 
fine frequency resolution (although coarse obviously). In 
Fig.~\ref{fig:timefreq}(a), for 
the experiment at 
$Re_f=300$, the subharmonic energy bumps start to be detectable after 
typically 100 forcing periods $T$. The amplitude of the subharmonic bumps then 
appear to 
slowly grow during the rest of the experiment. To have a more quantitative view,
we report in Fig.~\ref{fig:timefreq}(c) the time evolution of the energy 
density $E(\sigma,t,\Delta T)$ for four specific frequencies:
the forcing frequency $\sigma_0^*=0.84$, the subharmonic frequencies 
$\sigma_1^*$ and $\sigma_2^*$ (reported in Table~\ref{tab:freqb}) and the 
frequency $\sigma^*=\sigma_0^*/2$ which is shown as a tracer of the 
spectral noise level (see Fig.~\ref{fig:spkfreq}).
In Fig.~\ref{fig:timefreq}(c), we observe that the initial increase of the 
amplitude of the primary wave typically takes $15$ forcing 
periods $T$. Using the theoretical value of the group velocity $|{\bf 
c_g}|=2\Omega \sin \theta/k_0 \simeq 2.47$~cm/s of the primary wave, we can 
estimate 
the duration of the initial propagation of the primary wave through 
the studied region of $10 \times 12$~cm$^2$ area to be of about $2$ to $3$ 
forcing periods $T$. The apparent duration of $15\,T$ 
of the growth in 
amplitude of the forced wave in Fig.~\ref{fig:timefreq}(c) results from the 
time width of the 
sliding window used for the computation of the time-frequency spectra: 
processes taking place over a duration much shorter than $\Delta T$ have their 
duration artificially increased up to typically $\Delta T = 15\,T$. This 
artificial spreading in time also explains the fact that we already 
observe the growth of the amplitude of the primary wave at $t=0$ while it is 
expected to start only after about $6\,T$ 
after 
the start of the forcing (the dashed rectangle is at a distance of about 
$30$~cm from the wave maker).

In Fig.~\ref{fig:timefreq}(c), the amplitudes 
of the two subharmonic waves, which have a similar behavior, emerge from the 
spectral noise level around $t=100\,T$ after the start of the forcing, before 
they slowly grow until 
the end of the experiment at $t=300\,T$. By fitting the increase with time 
of the amplitude of the subharmonic bumps with the exponential behavior $E 
\sim \exp(2\gamma^{\rm (exp)} t)$ over the time period $100\,T<t<200\,T$, we 
estimate an experimental growth rate (for the velocity) of the 
subharmonic modes of $\gamma^{\rm (exp)} \simeq 3.1\,10^{-3}$~s$^{-1}$. This 
value 
is about fifty times smaller than the theoretical growth rate, of about 
$\gamma^{\rm (max)}\simeq 0.17$~s$^{-1}$, computed in the theoretical section~\ref{sec:theory} for a 
primary wave with features matching those of the experiment at 
$Re_f=300$ (details will be given in section~\ref{sec:wavevec} regarding this 
point). This discrepancy most 
probably reveals that the exponential growth of the subharmonic waves predicted 
at the onset of the TRI is restricted to earlier times in the experiment, 
before $t<100\,T$, for which unfortunately the subharmonic waves amplitude is 
too weak to be resolved by the PIV measurements. A natural interpretation for the low 
growth rate observed here for the subharmonic modes over the time period $100\,T<t<200\,T$ is that the saturation processes are already in action even if the 
saturation is not yet completed. Also, it is worth to note that we do not 
observe a 
significant broadening of the subharmonic bumps during their observable growth 
phase. This implies that the spreading in frequency of the subharmonic modes 
produced by the TRI does not necessarily result from the saturation 
processes of the instability.

For the experiment at $Re_f=420$ reported in Figs.~\ref{fig:timefreq}(b) 
and (d), 
the scenario of the instability takes place much faster: the 
subharmonic bumps become detectable beyond $t\simeq 45\,T$ and then 
saturate in amplitude around $t\simeq 75\,T$. The experimental growth rate 
for the velocity amplitude of the modes at $\sigma_1^*$ and $\sigma_2^*$ 
during the period $45\,T<t<75\,T$ is here of about $\gamma^{\rm (exp)} \simeq 
0.030$~s$^{-1}$. This value, much larger than the one measured for the 
experiment 
at $Re_f=300$, is still significantly smaller (seven times smaller) than the maximum 
theoretical 
growth rate, of about $\gamma^{\rm (max)}\simeq 0.20$~s$^{-1}$, that can be 
computed theoretically for a 
primary wave with features matching the ones of the experiment at 
$Re_f=420$. The observable growth of the subharmonic modes proceeds over a relatively short duration of about $30\,T$, of the same order as the width of 
the sliding window used to compute the spectra.
Contrary to the experiment at $Re_f=300$, it is therefore here most likely 
that the measured growth rate is significantly biased (reduced) by the temporal 
spectrum computation and it is possible that the actual growth rate is not that 
far from the theoretical one. A final remark worth to be 
done regarding Fig.~\ref{fig:timefreq} is the fact that the 
forced wave, after its initial propagation through 
the studied area, 
experiences a slow decrease in amplitude during the stage where 
the secondary waves are increasing in amplitude, before finally reaching a 
stable state when the secondary waves saturate.

\subsection{Comparison of the experimental data to the 
theory}\label{sec:wavevec}

In this section, we focus on the region of the flow where the instability develops (the dashed 
rectangle in Fig.~\ref{fig:vy_H_lb7p6_A1p5mm_O18rpm}), where the three waves at 
$\sigma_0^*$, $\sigma_1^*$ and $\sigma_2^*$ are all energetic. We measure for 
each subharmonic mode the wavevector components in the measurement plane, 
$k_{x,i}^{\rm exp}$ and $k_{z,i}^{\rm exp}$, by spatially averaging the phase 
field gradient $\nabla \varphi$ over that region (the dashed rectangle). The 
corresponding 
measurement errors are computed as the standard deviation of the phase 
field gradient $\nabla \varphi$ over the same area.

Considering the 
observations of the previous section, the subharmonic waves are propagating out 
of the measurement plane ($x,z$) and should therefore have a non-zero 
wavevector component along the $y$-direction. We estimate this component 
by means of the dispersion relation of inertial waves~(\ref{eq:dispersion}) as
\begin{equation}\label{eq:estky}
     k_{y,i}^{\rm exp}=\pm \sqrt{\left(k_{z, i}^{\rm exp}\right)^2\left(\frac{1}{{\sigma_i^*}^2}-1\right)-\left(k_{x, i}^{\rm exp}\right)^2} \,.
\end{equation}
For the secondary waves, we estimate $k_{y,i}^{\rm exp}$ by spatially averaging 
(\ref{eq:estky}) over the 
instability region (the dashed rectangle in 
Fig.~\ref{fig:vy_H_lb7p6_A1p5mm_O18rpm}). 
We 
compute the corresponding error as the standard deviation of (\ref{eq:estky}) 
over the same region.
As a test, we apply this procedure to the primary wave of the experiment at 
$Re_f=300$ which leads to a negligible out-of-plane wavevector component 
$|k_{y,0}^{\rm exp}|=0.017$~rad/cm (computed as the root of the spatial mean of 
$|{k_{y,0}^{\rm exp}}|^2$ over the instability region) and a wavenumber 
$k_0^{\rm exp}=0.83 \pm 0.02$~rad/cm in excellent agreement with the expected 
value 
$2\pi/(7.6$~cm$)\simeq 0.83$~rad/cm. The computed values of the wavevector 
components for the waves at $\sigma_1^*$ and $\sigma_2^*$ with the 
corresponding errors are reported in Table~\ref{tab:freqb} for the 
experiments at $Re_f=300$ and $Re_f=420$, for which the TRI is observed.

\begin{figure}
	\centerline{\includegraphics[width=14cm]{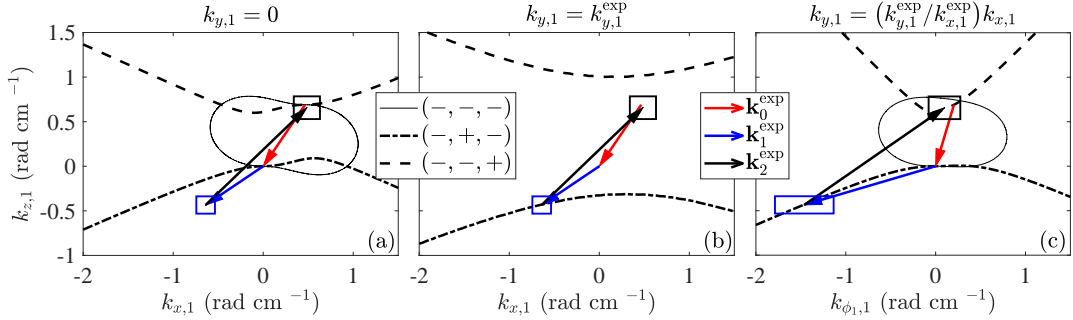}}
	\caption{Cuts of the theoretical resonance surfaces for ${\bf k_1}$ 
	computed for a primary wave of frequency $\sigma_0^*=0.84$, wavelength 
	$\lambda_f=7.6~$cm and amplitude $b_0=3.9$~mm/s matching the features of 
	the experimental primary wave at $Re_f=300$  in its unstable region (dashed 
	rectangle in Fig.~\ref{fig:vy_H_lb7p6_A1p5mm_O18rpm}). Superimposed on the 
	resonance curves, we show the projection on the considered plane of the 
	wavevectors ${\bf k_0^{\rm exp}}$, ${\bf k_1^{\rm exp}}$ and ${\bf k_2^{\rm 
	exp}}$ experimentally estimated for the experiment at $Re_f=300$. We 
	also report the measurement error corresponding to each wavevector 
	component via a rectangle around each wavevector tip.
	(a) Cut in the vertical plane $k_{y,1}=0$. (b) Cut in the vertical plane 
	$k_{y,1}=k_{y,1}^{\rm exp}$. (c) Cut in the vertical plane 
	$k_{y,1}/k_{y,1}^{\rm exp}=k_{x,1}/k_{x,1}^{\rm exp}$ 
	($k_{\phi_1,1}=\sqrt{k_{x,1}^2+k_{y,1}^2}=k_{x,1}\sqrt{1+(k_{y,1}^{\rm 
	exp}/k_{x,1}^{\rm exp})^2}).$}\label{fig:triads-case1}
\end{figure}

In Fig.~\ref{fig:triads-case1}, we report three different cuts of the resonance 
surfaces for ${\bf k_1}$ computed theoretically for a primary wave of frequency 
$\sigma_0^*=0.84$, wavelength $\lambda_f=7.6~$cm, and amplitude $b_0=3.9$~mm/s. 
These values match the features of the experimental primary wave at $Re_f=300$ 
in its unstable region (dashed rectangle in 
Fig.~\ref{fig:vy_H_lb7p6_A1p5mm_O18rpm}). For each cutting plane, the resulting 
curves for the three possible polarities combinations are shown. In panel (a), 
we report the classical resonance curves for ${\bf k_1}$ in the plane 
($k_x,k_z$) of the primary wave ($k_{y,1}=0$). We superimpose on this figure 
the projections of the experimentally computed wavevectors ${\bf k_1^{\rm 
exp}}$ and ${\bf k_2^{\rm exp}}$ and of the primary wavevector ${\bf k_0^{\rm 
exp}}$ on the ($k_x,k_z$) plane. We also report the measurement error 
of each wavevector component via a rectangle around each 
wavevector tip. We recall that the frequencies $\sigma_1$ and $\sigma_2$ are 
predicted negative by the theory (see 
Sec.~\ref{sec:theory}). In parallel, they are definite positive in the 
experimental data processing. In order to compare the experimental data to the 
theory, in the following, we therefore systematically multiply by $-1$ the experimentally 
measured wavevectors ${\bf k_1^{\rm exp}}$ and ${\bf k_2^{\rm exp}}$ before 
superimposing them on the theoretical curves, starting with 
Fig.~\ref{fig:triads-case1} [note that waves with ($\sigma,{\bf k}$) and 
($-\sigma,-{\bf k}$) are the same]. In 
Fig.~\ref{fig:triads-case1}(a), we observe that the three wavevectors form an 
almost closed triangle in the vertical plane ($k_x,k_z$), i.e., $ 
\mathbf{k}_0^{\rm exp}+\mathbf{k}_1^{\rm exp}+\mathbf{k}_2^{\rm exp}\simeq {\bf 
0}$: this confirms the spatial resonance of the three waves involved in the 
instability, at least in the vertical plane ($k_x,k_z$). Nevertheless, it is 
clear that the tip of the experimentally measured wavevector $(k_{x,1}^{\rm 
exp},k_{z,1}^{\rm exp})$ does not fall on one of the ${\bf k_1}$ resonance 
curves of the in-plane triadic resonance instability. In panel (b), we show the 
theoretical resonance curves for ${\bf k_1}$ in the plane $k_{y,1}=k_{y,1}^{\rm 
exp}$ on which we again superimpose the projection of the experimentally 
measured wavevectors ${\bf k_0^{\rm exp}}$, ${\bf k_1^{\rm exp}}$ and ${\bf 
k_2^{\rm exp}}$ on the ($k_x,k_z$) plane. This time the nearly closed 
wavevectors triad has its ${\bf k_1^{\rm exp}}$ tip almost exactly on the 
($-,+,-$) resonance curve. In the last panel (c), we show the theoretical ${\bf 
k_1}$ resonance curves in the plane $k_{y,1}/k_{y,1}^{\rm 
exp}=k_{x,1}/k_{x,1}^{\rm exp}$. In this plane, the projection of the 
experimental wavevectors triad is again almost closed and its ${\bf k_1}$ tip 
lies almost exactly on the ($-,+,-$) resonance curve. Altogether 
Fig.~\ref{fig:triads-case1} confirms, in agreement with the theoretical 
arguments of Sec.~\ref{sec:theory}, the experimental observation of a 
three-dimensional triadic resonance instability of type ($-,+,-$) driving the 
primary wave energy toward two subharmonic waves not propagating in the same 
vertical plane as the primary wave.

\begin{figure}
	\centerline{\includegraphics[width=7cm]{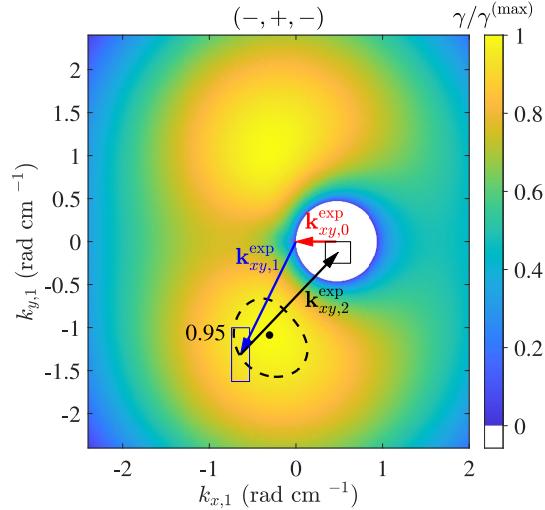}}
	\caption{Map of the growth rate $\gamma$ normalized by its maximum 
	$\gamma^{\rm (max)}$ as a function of $(k_{x,1},k_{y,1})$ for the ($-,+,-$) 
	instability computed theoretically for a primary wave with features 
	[$\sigma_0^*=0.84$, $\lambda_f=7.6~$cm, $b_0=3.9$~mm/s] matching the 
	experimental primary wave at $Re_f=300$ (same map as in 
	Fig.~\ref{fig:kz1_gtheo}). Superimposed on the $\gamma/\gamma^{\rm (max)}$ 
	map, we show the projection in the $(k_{x},k_{y})$ plane of the 
	experimental wavevectors 
	${\bf k_0^{\rm exp}}$, ${\bf k_1^{\rm exp}}$ and  ${\bf k_2^{\rm exp}}$.
	As in Fig.~\ref{fig:triads-case1}, the measurement errors on the 
	wavevectors are reported via a rectangle around each wavevector 
	tip.}\label{fig:gammakp-1} 
\end{figure}

To further compare the experimental data with the theoretical predictions, we 
report in Fig.~\ref{fig:gammakp-1} the map of the growth rate $\gamma$ as a 
function of $(k_{x,1},k_{y,1})$ for the ($-,+,-$) instability. This map is 
computed for a primary wave with features [$\sigma_0^*=0.84$, 
$\lambda_f=7.6~$cm, $b_0=3.9$~mm/s] matching the experimental primary wave 
characteristics at $Re_f=300$ (same parameters as in Fig.~\ref{fig:kz1_gtheo}). 
We superimpose to the map of $\gamma$ the projection on the $(k_{x},k_{y})$ 
plane of the experimental wavevectors ${\bf k_0^{\rm exp}}$, ${\bf k_1^{\rm 
exp}}$ and ${\bf k_2^{\rm exp}}$ at $Re_f=300$. First, we observe that the 
experimental triad in the ($k_x,k_y$) plane is also close to spatial resonance 
with the wavevectors tending to form a closed triangle. Moreover, the tip of 
the ${\bf k_1^{\rm exp}}$ wavevector is near the location of the maximum of the 
theoretical instability growth rate: it is included in the region where 
$\gamma$ is larger 
than $95\%$ of its maximum. The latter observation shows that the features of 
the triadic resonance instability observed in the experiment at $Re_f=300$ are 
consistent with the theoretical predictions based on the selection of the 
maximum growth rate. 

We recall here that the theory presented in Sec.~\ref{sec:theory} focuses 
on the early times of the instability during which the amplitudes of the 
subharmonic secondary waves grow exponentially from low values. In parallel, 
as 
already discussed, we are not able to study experimentally the initial 
exponential growth of the subharmonic waves because during this stage the 
subharmonic waves amplitude is smaller than the resolution of our PIV 
measurements. Therefore, the experimental characterization of the subharmonic 
waves is conducted over the whole experiment duration. The growing 
influence of the saturation processes 
observed during the experiment at $Re_f=300$ can possibly 
lead to discrepancies between the maximum growth rate modes expected to be 
dominant at the early stages of the instability and the dominant modes present 
in the experimentally studied stage where the subharmonic waves are detectable. 
It is therefore even more remarkable to 
observe such an excellent 
agreement between the theoretical predictions of the 3D TRI at early times and 
our measurements. Finally, recalling that the map of the theoretical growth 
rate 
$\gamma(k_{x,1},k_{y,1})$ is symmetric with respect to the $k_{y,1}=0$ axis 
with two symmetric maxima, we should mention that in Fig.~\ref{fig:gammakp-1} 
when computing $k_{y,1}^{\rm exp}$ from Eq.~(\ref{eq:estky}), we arbitrarily 
choose the sign of the $y$ wavevector component, the most 
probable situation being that both signs are present in reality. Another remark 
is worth to be done: References~\cite{Bourget2014} and \cite{Karimi2014} have 
shown that refinements to the model of the triadic resonance instability can be 
done in order to account for the finite size of the wave beam, i.e., for the 
finite number of wavelengths contained in the beam width. In 
Appendix~\ref{app:fis}, we show that, after including to the theory the finite size 
corrections proposed by Bourget~\textit{et al.}~\cite{Bourget2014}, 
the map of the theoretical growth rate is only slightly modified by the finite 
size effects 
and that the good agreement between the experimental triad and the 
most unstable theoretical triad is preserved.

\begin{figure}
	\centerline{\includegraphics[width=14cm]{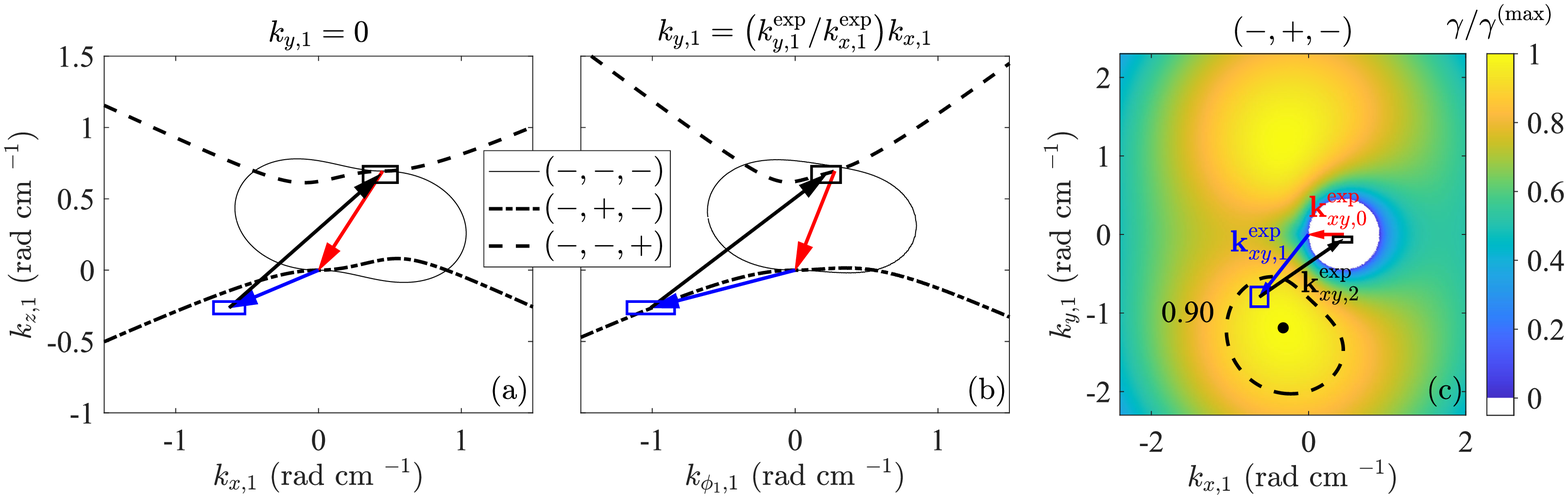}}
	\caption{(a-b) Cuts of the theoretical resonance surfaces for ${\bf k_1}$ computed for a primary wave of frequency $\sigma_0^*=0.84$, wavelength $\lambda_f=7.6~$cm and amplitude $b_0=5.1$~mm/s matching the features of the experimental primary wave at $Re_f=420$. Superimposed on the resonance curves, we show the projection on the considered plane of the wavevectors ${\bf k_0^{\rm exp}}$, ${\bf k_1^{\rm exp}}$ and ${\bf k_2^{\rm exp}}$ measured for the experiment at $Re_f=420$. (a) Cut in the vertical plane $k_{y,1}=0$. (b) Cut in the vertical plane $k_{y,1}/k_{y,1}^{\rm exp}=k_{x,1}/k_{x,1}^{\rm exp}$. (c) 
	Map of the growth rate $\gamma$ normalized by its maximum $\gamma^{\rm (max)}$ as a function of $(k_{x,1},k_{y,1})$ for the ($-,+,-$) instability computed theoretically for a primary wave matching the features of the experimental primary wave at $Re_f=420$. Superimposed on the $\gamma/\gamma^{\rm (max)}$ map, we show the projection in the $(k_{x},k_{y})$ plane of the experimental wavevectors ${\bf k_0^{\rm exp}}$, ${\bf k_1^{\rm exp}}$ and  ${\bf k_2^{\rm exp}}$.}\label{fig:triads-case2}
\end{figure}

In the following we reproduce the same analysis for the experiment at 
$Re_f=420$. In Fig.~\ref{fig:triads-case2}, we report the cut of the 
theoretical resonance surface for ${\bf k_1}$ in the planes $k_{y,1}=0$ (panel 
a) and  $k_{y,1}/k_{y,1}^{\rm exp}=k_{x,1}/k_{x,1}^{\rm exp}$ (panel b) on 
which we superimpose the projection of the experimental 
wavevectors triad corresponding to the maximum of energy in the first 
subharmonic bump 
of the temporal energy spectrum (Fig.~\ref{fig:spkfreq}). In 
Fig.~\ref{fig:triads-case2}(c), we show the 
map of the theoretical growth rate $\gamma$ as a function of 
$(k_{x,1},k_{y,1})$ for the ($-,+,-$) instability. This map is computed for a 
primary wave with features [$\sigma_0^*=0.84$, $\lambda_f=7.6~$cm, 
$b_0=5.1$~mm/s] matching the experimental primary wave characteristics at 
$Re_f\simeq 420$. We also superimpose to the map of $\gamma$ the projection on 
the $(k_{x},k_{y})$ plane of the experimental wavevectors triad at $Re_f=420$. 
In all panels of Fig.~\ref{fig:triads-case2}, we verify the fact that the 
experimental triad is nearly closed confirming the spatial resonance of the 
temporally resonant waves. In panel (b), the tip of the ${\bf k_1}$ 
experimental wavevector again falls very well on the $(-,+,-)$ theoretical 
resonance curve whereas it is at a significant distance from the 2D resonance 
curve in the $k_{y,1}=0$ plane (panel a). Finally, in panel (c), we observe 
that the tip of the ${\bf k_1}$ experimental wavevector is significantly remote 
from the maximum of the growth rate map. It is however found within the region 
where the theoretical growth rate $\gamma$ is larger than $90\%$ of its maximum 
which is rather satisfactory considering that, for this experiment at 
$Re_f=420$, the analysis is dominated by the saturated regime of the TRI which is not the case for the previously studied 
experiment at $Re_f=300$ which is closer to the instability onset.

\section{Conclusion}

In this article, we report PIV measurements of the velocity field produced in a 
rotating fluid by a wave generator. The wave maker is designed to produce a 
wave beam approaching the structure of a plane inertial wave. In practice, the 
wave beam contains four wavelengths in its width. Besides, the wave generator 
is particularly large in the horizontal direction $y$ in which the plane wave is supposed to be invariant: the generator extension in the 
$y$-direction corresponds to nearly seven wavelengths. This last feature is 
radically
different from previous experiments aiming to produce plane inertial (or 
internal gravity) waves where the wave maker extension in the $y$-direction was 
(slightly) smaller than two 
wavelengths~\cite{Bordes2012,Joubaud2012,Bourget2013,Bourget2014}. Starting 
from the linear regime, we increase the forcing amplitude in order to explore 
the emergence of the non-linear effects affecting the forced inertial wave. 
Above a given threshold in amplitude, the forced wave is subject to 
an instability transferring some of its energy toward two subharmonic inertial 
waves in temporal and spatial triadic resonance with the primary wave. We 
nevertheless show that the secondary waves are not propagating in the same 
vertical plane as the primary wave: they are non-invariant in the horizontal 
direction $y$ along which the primary wave is invariant. This spontaneous 
breaking of the symmetry of the base flow shows that the triadic resonance 
instability of the forced inertial wave is three-dimensional.

In parallel, by building on the classical inertial wave triadic interaction 
coefficients, we compute numerically the growth rate of the triadic resonance 
instability (TRI) of a plane inertial wave in the three-dimensional case. We 
show that the maximum growth rate is associated with a three-dimensional 
instability producing two secondary waves propagating out of the primary wave 
vertical plane. We also show that this result can be demonstrated analytically 
in the inviscid case where the TRI becomes a Parametric Subharmonic Instability 
(PSI)~\cite{Staquet2002,Dauxois2018} with two secondary waves at vanishing scale and at frequencies equal to 
half the primary wave frequency. Finally, we demonstrate that the secondary 
wavevectors observed in our experiments agree well with the triad predicted 
theoretically by the maximization of the theoretical instability growth rate. 
This agreement with the theory for a plane wave confirms that the 
three-dimensionality of the TRI observed experimentally is intrinsic and 
unrelated to deviations of the experimental primary wave from an exact plane 
wave (due to friction on the water tank walls, finite size effects ...).  

An important consequence of our results concerns flows in the wave 
turbulence 
regime (at larger Reynolds number than the ones considered 
here)~\cite{Monsalve2020,LeReun2021}. One of the key assumptions made when 
deriving 
the scaling laws for the spatial energy spectrum from the kinetic equations in 
weak inertial-wave turbulence theory~\cite{Galtier2003} is the statistical 
axi\-symmetry of the flow around the rotation axis. We have shown in the 
present article that the triadic resonant interactions between inertial waves 
are very efficient at redistributing the energy in the horizontal plane, normal 
to rotation.
This feature should contribute to drive flows in the inertial wave turbulence 
regime toward statistical axisymmetry and to make them fulfill the assumption 
made in the derivation of the wave turbulence theory~\cite{Galtier2003}.

At this point, an interesting question concerns the triadic resonance 
instability of an internal gravity wave: is it two or three-dimensional ?
In the 2D case (invariant in horizontal direction $y$), the expression of the 
triadic interaction coefficients and of the growth rate of the TRI for internal 
gravity waves are analogous to those for inertial 
waves~\cite{Maurer2016}. However, this similarity seems not to hold when 
considering 3D triadic interactions of waves propagating in different vertical 
planes~\cite{Remmel2014} and only a dedicated study will provide answers in the 
case of internal gravity waves. The question of the 
three-dimensionality of the TRI for an internal gravity wave has been recently 
tackled by Ghaemsaidi and Mathur~\cite{Ghaemsaidi2019} who implemented 
a local stability analysis for a plane internal gravity wave. Their analysis, 
restricted to the inviscid limit and to small scale perturbations ---i.e., the 
Parametric Subharmonic Instability case--- shows that a three-dimensional PSI 
is possible for an internal gravity wave. However, the instability associated 
with the maximum growth rate is shown to remain two-dimensional with secondary 
waves propagating in the same vertical plane as the primary wave. The local 
stability analysis of Ghaemsaidi and Mathur predicts that the internal wave 
instability starts to be dominated by three-dimensional processes when the 
primary wave becomes strongly non-linear, i.e. with a Froude number (equivalent to 
the Rossby number in stratified fluids) larger than $1$. In this situation, the 
instability growth rate is shown to be larger than the internal wave 
frequencies: the flow is completely out of the weakly non-linear framework of 
the triadic resonance instability.

The three-dimensionality of the TRI of an internal gravity wave toward 
two subharmonic waves of finite wavelengths remains an open question, to be 
investigated theoretically and experimentally. More generally, the question of 
the energy redistribution in the horizontal plane normal to gravity by internal 
gravity wave triadic interactions remains open with important stakes regarding 
the conditions under which the wave turbulence formalism for stratified 
fluids~\cite{Caillol2000,Lvov2001,Lvov2004} could be relevant.

\begin{acknowledgments}
We acknowledge J. Amarni, A. Aubertin, L. Auffray and R. Pidoux for 
experimental help. This work was supported by grants from the Simons Foundation 
(651461 PPC and 651475 TD), and by the Agence Nationale de la Recherche through 
Grant ``DisET'' No.~ANR-17-CE30-0003.
\end{acknowledgments}

\appendix
\section{Asymptotic expression of the growth rate in the inviscid limit}
\label{app:gammaRe0inf}
Using Eq.~(\ref{eq:growthrate}), the growth rate of the triadic resonance instability of a plane inertial wave is given by
\begin{equation}
    \gamma=\frac{-\nu(k_1^2+k_2^2)}{2}+\sqrt{\frac{\nu^2(k_1^2-k_2^2)^2}{4}+C_1 \overline{C_2} |b_0|^2}\, ,
\end{equation}
where, according to Waleffe~\cite{Waleffe1992},
\begin{equation}
C_1 \overline{C_2}=\frac{\sin^2\alpha_2}{4k_2^2}\left(s_0k_0+s_1k_1+s_2k_2\right)^2\left(s_0k_0-s_2k_2\right)\left(s_1k_1 - s_0k_0\right)\,.
\end{equation} 

Following Eq.~(8) of Smith and Waleffe~\cite{Smith1999} (which is the direct 
consequence of the dispersion relation combined to the triadic resonance 
conditions~(\ref{eq:temp_res}-\ref{eq:spat_res})), one can show that
\begin{equation}
    \left(s_0k_0-s_2k_2\right)=\frac{\sigma_1}{\sigma_0}\left(s_2k_2-s_1k_1\right) \quad {\rm and} \quad \left(s_0k_0-s_1k_1\right)=\frac{\sigma_2}{\sigma_0}\left(s_1k_1-s_2k_2\right)\, ,
\end{equation}
such that
\begin{equation}
C_1 \overline{C_2}=\frac{\sin^2\alpha_2}{4k_2^2}\left(s_0k_0+s_1k_1+s_2k_2\right)^2\frac{\sigma_1\sigma_2}{\sigma_0^2}\left(s_1k_1-s_2k_2\right)^2\, .
\end{equation} 

In the following, we conduct an asymptotic expansion of the inviscid growth 
rate $\gamma = |b_0| \sqrt{C_1 \overline{C_2}}$ to the first order in $k_0/k_1 
\simeq k_0/k_2$ assuming that the secondary wavenumbers associated to the 
maximum growth rate are much larger than the primary wavenumber such that 
$|\sigma_1|=|\sigma_2|=\sigma_0/2$ and $k_1\simeq k_2 \gg k_0$. Focusing on the 
combination of wave polarities $(s_0=-1,s_1=+1,s_2=-1)$, the growth rate can be 
written
\begin{eqnarray}
    \gamma&\simeq &\frac{|b_0|}{4 k_2} \sin\alpha_2 \sqrt{\left(k_0-k_1+k_2\right)^2\left(k_2+k_1\right)^2}\, ,\\
          &\simeq &\frac{|b_0|}{4 k_2} \sin\alpha_2 \sqrt{\left(k_0(k_1+k_2)+k_2^2-k_1^2\right)^2}\, .\label{eq:gamma101}
\end{eqnarray}
Using the law of cosines
\begin{equation}\label{eq:cos}
    \cos\alpha_2=\frac{k_0^2+k_1^2-k_2^2}{2k_1k_0},
\end{equation}
Eq.~(\ref{eq:gamma101}) gives, to the first order in $k_0/k_1 \simeq k_0/k_2$,
\begin{eqnarray}
    \frac{\gamma}{|b_0|k_0}&\simeq &\frac{1}{2} \sin\alpha_2 (1-\cos\alpha_2)\, .\label{eq:gamma102}
\end{eqnarray}

Maximizing equation~(\ref{eq:gamma102}) with respect to $\alpha_2$ yields
\begin{equation}
    \cos\alpha_2=-1/2
\end{equation}
corresponding to an angle $\alpha_2 = 2\pi/3$~rad ($= 120^\circ$) and to a growth rate equal to 
\begin{equation}
    \gamma^{\rm (max)} \simeq 0.6495\, |b_0|k_0.
    \label{eq:asymg}
\end{equation}

Injecting this specific value $\alpha_2 = 2\pi/3$~rad in the law~(\ref{eq:cos}), one gets
\begin{equation}\label{eq:cos2}
    k_2^2=k_1^2+k_0^2+k_0k_1 \, .
\end{equation}
Retaining the sign conventions used in Sec.~\ref{sec:theory}, the primary wave vector can be written ${\bf k_0}=(k_{x,0},0,k_{z,0})=(-k_0 \sin\theta_0,0, - k_0\cos\theta_0)$ with $\theta_0=\cos^{-1}(\sigma_0^*)$. Besides, we note ${\bf k_1}=(k_{x,1},k_{y,1},k_{z,1})=(-k_1 \sin\theta_1 \cos\phi_1,-k_1 \sin\theta_1 \sin\phi_1, -k_1\cos\theta_1)$ the components of the wave vector of secondary wave $1$ with $\theta_1=\cos^{-1}(|\sigma_1^*|)$. 
Using the spatial resonance condition
\begin{equation}
    k_2^2=(k_{x,1}+k_{x,0})^2+k_{y,1}^2+(k_{z,1}+k_{z,0})^2\, ,
\end{equation}
one gets
\begin{eqnarray}
    k_0\,k_1 &=& 2\,{\bf k_0}\cdot{\bf k_1}\, ,\\
             &=&   k_0\,k_1 \left(2\sin\theta_0\sin\theta_1\cos\phi_1+2\cos\theta_0\cos\theta_1\right)\, . 
\end{eqnarray}
Using now the fact that the secondary wave $1$ maximizing the growth rate verifies $|\sigma_1^*|=\cos\theta_1=\sigma_0^*/2=\cos\theta_0/2$ in the inviscid limit, we obtain
\begin{eqnarray}
\sin\theta_0=2\sin\theta_1\cos\phi_1\, ,
\end{eqnarray}
which finally leads to
\begin{equation}
    \cos^2\phi_1=\frac{1-{\sigma^*_0}^2}{4-{\sigma^*_0}^2}\, ,
\end{equation}
and to 
\begin{equation}
   \frac{k_{y,1}}{k_{x,1}}=\tan\phi_1=\pm \sqrt{\frac{3}{1-{\sigma^*_0}^2}}\,.
\end{equation}

\section{Accounting for finite size effects}\label{app:fis}

In Ref.~\cite{Bourget2014}, Bourget~\textit{et al.} propose a refined 
theoretical description of the triadic resonance instability of an internal 
gravity wave accounting for the finite size of the wave beam (see also 
Ref.~\cite{Karimi2014} on this topic). They consider the instability of a 
monochromatic wave beam modulated in its transverse direction by a rectangular 
function of width $W$. Their description is based on an energy budget realized 
inside a control volume matching the beam width $W$ where the TRI takes place. 
In the following, we reproduce this model for an inertial wave and modify it 
accordingly: the viscous dissipation rate for an inertial wave of wavenumber 
$k$ is $\nu k^2$~\cite{Machicoane2018} whereas it is $\nu k^2/2$ for an internal gravity wave~\cite{Sutherland2010}. Then, 
in the equations of evolution of the secondary wave amplitudes 
(Eqs.~\ref{eq:b1}-\ref{eq:b2}), the dissipation rate $\nu k_i^2$ for the 
secondary wave $i$ must be replaced by $\chi_i=\nu k_i^2+ |{\bf c_{g,i}}\cdot 
{\bf{\hat{k}_0}}|/2W$ with ${\bf c_{g,i}}$ the group velocity of wave $i$ and 
$\bf{\hat{k}_0}$ the unit vector parallel to the primary wave vector. The 
additional term $|{\bf c_{g,i}}\cdot {\bf{\hat{k}_0}}|/2W$ actually accounts 
for the rate at which the energy of the secondary wave $i$ is leaving the 
control volume where the instability takes place. Finally, the TRI growth 
rate for an inertial wave beam of transverse width $W$ becomes
\begin{equation}\label{eq:FSE}
    \gamma=-\frac{\chi_1+\chi_2}{2}+\sqrt{\frac{(\chi_1-\chi_2)^2}{4}+\textcolor{black}{C_1 \overline{C_2}} |b_0| ^2}\, .
\end{equation}
The contribution of the additional damping term $|{\bf c_{g,i}}\cdot 
{\bf{\hat{k}_0}}|/2W$ due to the beam finite size will reduce the growth rate 
and stabilize certain triads that were unstable for an ``infinite'' plane wave.

\begin{figure}
	\centerline{\includegraphics[width=13cm]{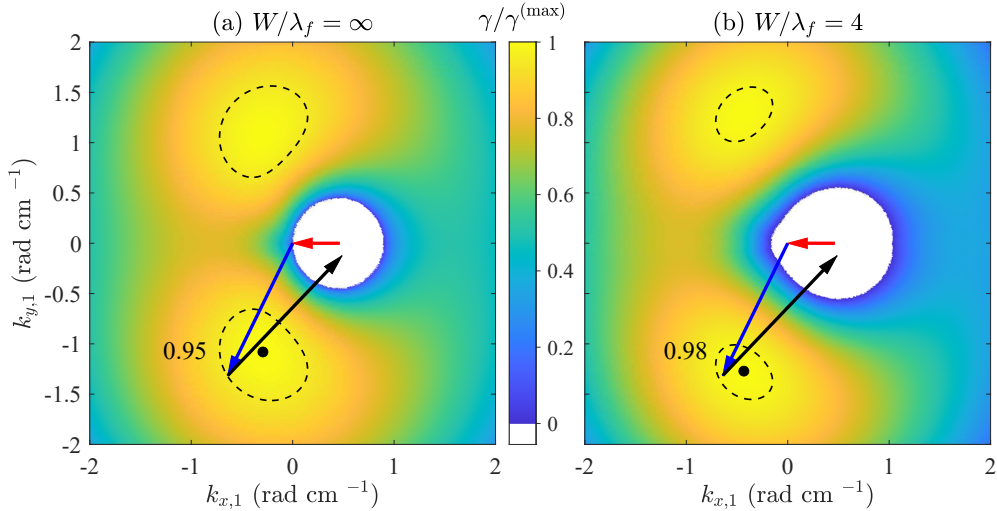}}
	\caption{Map of the growth rate $\gamma$ normalized by its maximum 
	$\gamma^{\rm (max)}$ as a function of $(k_{x,1},k_{y,1})$ for the ($-,+,-$) 
	instability computed theoretically for a primary wave with features 
	[$\sigma_0^*=0.84$, $\lambda_f=7.6~$cm, $b_0=3.9$~mm/s] matching the 
	experimental primary wave at $Re_f=300$. Superimposed on the 
	$\gamma/\gamma^{\rm (max)}$ map, we show the projection in the 
	$(k_{x},k_{y})$ plane of the wavevectors ${\bf k_0^{\rm exp}}$, ${\bf 
	k_1^{\rm exp}}$ and  ${\bf k_2^{\rm exp}}$ estimated experimentally. (a) 
	Growth rate map for an (infinite) plane wave (same map as in 
	Fig.~\ref{fig:gammakp-1}). (b) Growth rate map for a wave beam with four 
	wavelengths in its width $W=4\lambda_f$ following 
	Eq.~(\ref{eq:FSE}).}\label{fig:gammaannexe} 
\end{figure}

It is interesting to analyze to what extent including these finite size effects 
modifies the theoretical prediction for the most unstable triad, and its 
agreement with the experimental triad reported in the present article. Focusing 
on the primary wave parameters of the experiment at $Re_f=300$ (analyzed in 
Figs.~\ref{fig:vy_H_lb7p6_A1p5mm_O18rpm}-\ref{fig:gammakp-1}), we report in 
Fig.~\ref{fig:gammaannexe} the map of the theoretical growth rate for a primary 
plane wave on the left and for a wave beam of width $W=4\lambda_f$ on the 
right, corresponding to the experimental wave beam containing four wavelengths 
in its width. As in Fig.~\ref{fig:gammakp-1}, we superimpose on these maps the 
experimental wavevectors triad projected on the plane 
($k_{x,1},k_{y,1}$) (Fig.~\ref{fig:gammaannexe}(a) is identical to 
Fig.~\ref{fig:gammakp-1}).

Overall, the map of the growth rate is only slightly modified by the finite 
size effects. For instance, in the plane ($k_{x,1},k_{y,1}$) the tip of the 
theoretical wavevector ${\bf k_1}$ associated with the maximum growth rate is 
shifted of a relative distance of $20\%$ (compared to the value of 
$\sqrt{k_{x,1}^2+k_{y,1}^2}$). Besides, there is a better agreement between the 
experimental wavevectors triad and the theoretical triad with the 
maximum growth rate: the tip of the experimental vector ${\bf k_1}$ is now 
included in the region where the theoretical growth rate is larger than $98\%$ 
of its maximum whereas it was $95\%$ when finite size effects where not 
accounted for. We realize that the conclusions put forward in the present 
article are not modified by the finite size of the primary wave beam: the 
refined model does still predict that the TRI is tridimensional and the 
agreement between the experimental TRI and the theory is conserved (and even 
slightly improved).


\begin{thebibliography}{99}

\bibitem{Greenspan1968} H.P. Greenspan, \textit{The Theory of Rotating Fluids} (Cambridge University Press, Cambridge, UK, 1968).

\bibitem{Lighthill1978} J. Lighthill, \textit{Waves in Fluids} (Cambridge University Press, Cambridge, UK, 1978).

\bibitem{Sutherland2010} B.R. Sutherland, \textit{Internal Gravity Waves} (Cambridge University Press, Cambridge, UK, 2010).

\bibitem{Brunet2019} M. Brunet, T. Dauxois, and P.-P. Cortet, Linear and non linear regimes of an inertial wave attractor, Phys. Rev. Fluids. \textbf{4}, 034801 (2019). 

\bibitem{Mowbray1967} D.E. Mowbray and B.S.H. Rarity, A theoretical and experimental investigation of the phase configuration of internal waves of small amplitude in a density stratified liquid, J. Fluid Mech. \textbf{28}, 1 (1967).
\bibitem{Thomas1972} N.H. Thomas and T.N. Stevenson, A similarity solution 
for viscous internal waves, J. Fluid Mech. \textbf{54}, 495 (1972).

\bibitem{Flynn2003} M. R. Flynn, K. Onu, and B. R. Sutherland, Internal wave excitation by a vertically oscillating sphere, J. Fluid Mech. \textbf{494}, 65 (2003).

\bibitem{Cortet2010} P.-P. Cortet, C. Lamriben, and F. Moisy, Viscous spreading of an inertial wave beam in a rotating fluid, Phys. Fluids \textbf{22}, 086603 (2010).

\bibitem{Machicoane2015} N. Machicoane, P.-P. Cortet, B. Voisin, and F. Moisy, Influence of the multipole order of the source on the decay of an inertial wave beam in a rotating fluid, Phys. Fluids \textbf{27}, 066602 (2015).

\bibitem{Mercier2010} M. Mercier, D. Martinand, M. Mathur, L. Gostiaux, T. Peacock, and T. Dauxois, New wave generation, J. Fluid Mech. \textbf{657}, 310 (2010).

\bibitem{Bordes2012} G. Bordes, F. Moisy, T. Dauxois, and P.-P. Cortet, Experimental evidence of a triadic resonance of plane inertial waves in a rotating fluid, Phys. Fluids \textbf{24}, 014105 (2012).

\bibitem{Bourget2013} B. Bourget, T. Dauxois, S. Joubaud and P. Odier, Experimental study of parametric subharmonic instability for internal plane waves, J. Fluid Mech. \textbf{723}, 1 (2013).

\bibitem{Aldridge1969} K.D. Aldridge and A. Toomre, Axisymmetric inertial oscillations of a fluid in a rotating spherical container, J. Fluid Mech. \textbf{37}, 307 (1969).

\bibitem{McEwan1970} A.D. McEwan, Inertial oscillations in a rotating fluid cylinder, J. Fluid Mech. \textbf{40}, 603 (1970).

\bibitem{Maas2003b} L.R.M. Maas, On the amphidromic structure of inertial waves in rectangular parallelepiped, Fluid Dyn. Res. \textbf{33}, 373 (2003).

\bibitem{Boisson2012b} J. Boisson, C. Lamriben, L.R.M. Maas, P.-P. Cortet, and F. Moisy, Inertial waves and modes excited by the libration of a rotating cube, Phys. Fluids \textbf{24}, 076602 (2012).

\bibitem{Boisson2012} J. Boisson, D. C\'ebron, F. Moisy, and P.-P. Cortet, Earth rotation prevents exact solid-body rotation of fluids in the laboratory, EPL \textbf{98}, 59002 (2012).

\bibitem{Maas1997} L.R.M. Maas, D. Benielli, J. Sommeria, and F.-P. A. Lam, Observation of an internal wave attractor in a confined, stably stratified fluid, Nature \textbf{388}, 557 (1997).

\bibitem{Rieutord2001} M. Rieutord, B. Georgeot, and L. Valdettaro, Inertial waves in a rotating spherical shell: attractors and asymptotic spectrum, J. Fluid Mech. \textbf{435}, 103 (2001).

\bibitem{Manders2003} A.M.M. Manders and L.R.M. Maas, Observations of inertial 
waves in a rectangular basin with one sloping boundary, J. Fluid Mech. {\bf 
493}, 39 (2003).

\bibitem{Grisouard2008} N. Grisouard, C. Staquet, and I. Pairaud, Numerical simulation of a two-dimensional internal wave attractor, J. Fluid Mech. \textbf{614}, 1 (2008).

\bibitem{Klein2014} M. Klein, T. Seelig, M.V. Kurgansky, A. Ghasemi, I.D. Borcia, A. Will, E. Schaller, C. Egbers and U. Harlander, Inertial wave excitation and focusing in a liquid bounded by a frustum and a cylinder, J. 
Fluid Mech. \textbf{751}, 255 (2014).

\bibitem{Pedlosky1987} J. Pedlosky, \textit{Geophysical Fluid Dynamics} (Springer-Verlag, New York, 1987).

\bibitem{Zakharov1992} V.E. Zakharov, V.S. L'vov, and G. Falkovich, {\it Kolmogorov Spectra of Turbulence} (Springer, Berlin, 1992).
	
\bibitem{Newell2011} A.C. Newell and B. Rumpf, Wave Turbulence, Annu. Rev. Fluid Mech. \textbf{43}, 59 (2011).

\bibitem{Nazarenko2011} S. Nazarenko, {\it Wave  Turbulence} (Springer, Berlin, 2011).

\bibitem{Gregg2018} M.C. Gregg, E.A. D'Asaro, J.J. Riley, and E. Kunze, Mixing efficiency in the ocean, Ann. Rev. Mar. Sci. \textbf{10}, 443 (2018).

\bibitem{Monsalve2020} E. Monsalve, M. Brunet, B. Gallet, P.-P. Cortet, Quantitative Experimental Observation of Weak Inertial-Wave Turbulence, Phys. Rev. Lett. \textbf{125}, 254502 (2020).

\bibitem{Yokoyama2020} N. Yokoyama and M. Takaoka, Energy-flux vector in anisotropic turbulence: Application to rotating turbulence, J. Fluid Mech. \textbf{908}, A17 (2021).

\bibitem{LeReun2021} T. Le Reun, B. Favier, and M. Le Bar, Evidence of the Zakharov-Kolmogorov spectrum in numerical simulations of inertial wave turbulence, Europhys. Lett. \textbf{132}, 64002 (2020).

\bibitem{Savaro2020} C. Savaro, A. Campagne, M. Calpe Linares, P. Augier, J. Sommeria, T. Valran, S. Viboud, and N. Mordant, Generation of weakly nonlinear turbulence of internal gravity waves in the Coriolis facility, Phys. Rev. Fluids \textbf{5}, 073801 (2020).

\bibitem{Davis2020} G. Davis, T. Jamin, J. Deleuze, S. Joubaud, and T. Dauxois, Succession of Resonances to Achieve Internal Wave Turbulence, Phys. Rev. Lett. \textbf{124}, 204502 (2020).

\bibitem{Galtier2003} S. Galtier, Weak inertial-wave turbulence theory, Phys. Rev. E \textbf{68}, 015301 (2003).

\bibitem{Cambon2004} C. Cambon, R. Rubinstein, and F.S. Godeferd, Advances in wave turbulence: rapidly rotating flows, New J. Phys. \textbf{6}, 73 (2004).

\bibitem{Lvov2004} Y.V. Lvov, K.L. Polzin, and E.G. Tabak, Energy Spectra of the Ocean's Internal Wave Field: Theory and Observations, Phys. Rev. Lett. \textbf{92}, 128501 (2004). 

\bibitem{Nazarenko2011b} S.V. Nazarenko and A.A. Schekochihin, Critical balance 
in magnetohydrodynamic, rotating and stratified turbulence: towards a universal 
scaling conjecture, J. Fluid Mech. \textbf{677}, 134 (2011).

\bibitem{Smith1999} L.M. Smith and F. Waleffe, Transfer of energy to 
two-dimensional large scales in forced, rotating three-dimensional turbulence, 
Phys. Fluids \textbf{11}, 1608 (1999).

\bibitem{Staquet2002} C. Staquet and J. Sommeria, Internal gravity waves: From instabilities to turbulence, Annu. Rev. Fluid Mech. \textbf{34}, 559 (2002).

\bibitem{Koudella2006} C. R. Koudella and C. Staquet, Instability mechanisms of a two-dimensional progressive internal gravity wave, J. Fluid Mech. \textbf{548}, 165 (2006).

\bibitem{Joubaud2012} S. Joubaud, J. Munroe, P. Odier, and T. Dauxois, Experimental parametric subharmonic instability in stratified fluids, Phys. Fluids \textbf{24}, 041703 (2012).

\bibitem{Bourget2014} B. Bourget, H. Scolan, T. Dauxois, M. Le Bars, P. Odier, and S. Joubaud, Finite-size effects in parametric subharmonic instability, J. Fluid Mech. \textbf{759}, 739 (2014).

\bibitem{Jouve2014} L. Jouve and G. I. Ogilvie, Direct numerical simulations of an inertial wave attractor in linear and nonlinear regimes, J. Fluid Mech. \textbf{745}, 223 (2014).

\bibitem{Scolan2013} H. Scolan, E. Ermanyuk, and T. Dauxois, Nonlinear fate of internal waves attractors, Phys. Rev. Letters \textbf{110}, 234501 (2013).

\bibitem{Brouzet2017} C. Brouzet, E. Ermanyuk, S. Joubaud, G. Pillet, and T. Dauxois, Internal wave attractors: different scenarios of instability, J. Fluid Mech. \textbf{811}, 544 (2017). 

\bibitem{Karimi2014} H.H. Karimi and T.R. Akylas, Parametric subharmonic 
instability of internal waves: Locally confined beams versus monochromatic wave 
trains, J. Fluid Mech. \textbf{757}, 381 (2014).

\bibitem{Machicoane2018} N. Machicoane, V. Labarre, B. Voisin, F. Moisy, 
and P.-P. Cortet, Wake of inertial waves of a horizontal cylinder in horizontal 
translation, Phys. Rev. Fluids \textbf{3}, 034801 (2018).

\bibitem{Lighthill1967} M.J. Lighthill, On waves generated in dispersive 
systems by travelling forcing effects, with applications to the dynamics of 
rotating fluids, J. Fluid Mech. \textbf{27}, 725 (1967).

\bibitem{LeReun2020} T. Le Reun, B. Gallet, B. Favier, and M. Le Bars, Near-resonant instability of geostrophic modes: beyond Greenspan's theorem, J. Fluid Mech. \textbf{900}, R2 (2020).

\bibitem{Brunet2020} M. Brunet, B. Gallet, and P.-P. Cortet, Shortcut to Geostrophy in Wave-Driven Rotating Turbulence: The Quartetic Instability, Phys. Rev. Lett. \textbf{124}, 124501 (2020). 

\bibitem{Craik1978} A.D.D. Craik and J.A. Adam, Evolution in space and time of resonant wave triads-I. The `pump-wave approximation', Proc. R. Soc. Lond. A. \textbf{363}, 243 (1978).

\bibitem{Gururaj2020} S. Gururaj and A. Guha, Energy transfer in resonant and near-resonant internal wave triads for weakly non-uniform stratifications. Part 1. Unbounded domain, J. Fluid Mech. \textbf{899}, A6 (2020).

\bibitem{Waleffe1992} F. Waleffe, The nature of triad interactions in homogeneous turbulence, Phys. Fluids A \textbf{4}, 350 (1992).

\bibitem{Waleffe1993} F. Waleffe, Inertial transfers in the helical decomposition, Phys. Fluids A \textbf{5}, 677 (1993).

\bibitem{Triana2012} S.A. Triana, D.S. Zimmerman, and D.P. Lathrop, Precessional states in a laboratory model of the Earth's core, J. Geophys. Res. \textbf{117}, B04103 (2012).

\bibitem{LeReun2019} T. Le Reun, B. Favier, and M. Le Bars, Experimental study of the nonlinear saturation of the elliptical instability: inertial wave turbulence versus geostrophic turbulence, J. Fluid Mech. \textbf{879}, 296 (2019).

\bibitem{Mercier2008} M.J. Mercier, N.B. Garnier, and T. Dauxois, Reflection 
and diffraction of internal waves analyzed with the Hilbert transform, Phys. 
Fluids \textbf{20}, 086601 (2008).

\bibitem{Dauxois2018} T. Dauxois, S. Joubaud, P. Odier, and A. Venaille, 
Instabilities of Internal Gravity Wave Beams, Annu. Rev. Fluid Mech. 
\textbf{50}, 131 (2018).

\bibitem{Maurer2016}
P. Maurer, S. Joubaud, and P. Odier, Generation and stability of inertial gravity waves, J. Fluid Mech. \textbf{808}, 539 (2016).

\bibitem{Remmel2014} M. Remmel, J. Sukhatme, and L.M. Smith, Nonlinear gravity-wave interactions in stratified turbulence, Theor. Comput. Fluid Dyn. \textbf{28}, 131 (2014).

\bibitem{Ghaemsaidi2019} S. Ghaemsaidi and M. Mathur, Three-dimensional small-scale instabilities of plane internal gravity waves, J. Fluid Mech. \textbf{863}, 702 (2019). 

\bibitem{Lvov2001} Y.V. Lvov and E.G. Tabak, Hamiltonian Formalism and the Garrett-Munk Spectrum of Internal Waves in the Ocean, Phys. Rev. Lett. \textbf{87}, 168501 (2001).

\bibitem{Caillol2000} P. Caillol and V. Zeitlin, Kinetic equations and stationary energy spectra of weakly nonlinear internal gravity waves, Dyn. Atmos. Oceans \textbf{32}, 81 (2000).

\end{thebibliography}
\end{document}